\documentclass[reprint,preprintnumbers,amsmath,aip,amssymb]{revtex4-1}
\usepackage{graphicx}% Include figure files
\usepackage{dcolumn}% Align table columns on decimal point
\usepackage{bm}% bold math
\usepackage{epsfig}
\usepackage{xcolor}
\usepackage{amsmath}
\usepackage{tikz}
\usetikzlibrary{matrix}
\usepackage{multirow}
\usepackage{ctable}

\graphicspath{{./EPS/}}

\def \be {\begin{equation}}
\def \ee {\end{equation}}
\def \ben {\begin{eqnarray}}
\def \een {\end{eqnarray}}
\def\re#1{{\color{black}{#1}}}
\begin{document}

%\preprint{APS/123-QED}

\title{Partially polaron-transformed quantum master equation for exciton and charge transport dynamics}
\author{Seogjoo J. Jang}
\email{seogjoo.jang@qc.cuny.edu}
\affiliation{Department of Chemistry and Biochemistry, Queens College, City University of New York, 65-30 Kissena Boulevard, Queens, New York 11367 \& PhD Programs in Chemistry and in Physics, Graduate Center of the City University of New York}  
\affiliation{Korea Institute for Advanced Study, Seoul 02455, South Korea}%Lines break automatically or can be forced with \\
%\\This line break forced with \textbackslash\textbackslash
%}%

\date{Published in {\it the Journal of Chemical Physics} {\bf 157}, 104107 (2022)}

\begin{abstract}
 Polaron-transformed quantum master equation (PQME) offers a unified framework to describe the dynamics of quantum systems in both limits of weak and strong couplings to environmental degrees of freedom. Thus, PQME serves as an efficient method to describe charge and exciton transfer/transport dynamics for a broad range of parameters in condensed or complex environments.   
However, in some cases, the polaron transformation (PT) being employed in the formulation invokes an over-relaxation of slow modes and results in premature suppression of important coherence terms.  A formal framework to address this issue \re{is developed in the present work }by employing a partial PT that has smaller weights for low frequency bath modes.  It is shown here that a closed form expression of a 2nd order time-local PQME including all the inhomogeneous terms can be derived for a general form of partial PT, although more complicated than that for the full PT.  All the expressions needed for numerical calculation are derived in detail.  Applications to a model \re{of two-level system coupled to a bath of harmonic oscillators}, with test calculations \re{focused on those due to homogeneous relaxation terms}, demonstrate the feasibility and the utility of the present approach.  
\end{abstract}

%\pacs{Valid PACS appear here}% PACS, the Physics and Astronomy
                             % Classification Scheme.
%\keywords{Suggested keywords}%Use showkeys class option if keyword
                              %display desired
\maketitle
\section{Introduction}
Polaron transformation (PT)\cite{landau-ujp53,frohlich-ap3,holstein-ap8-1,holstein-ap8-2,holstein-pmb37,emery-prb9,rackovsky-mp25,jackson-jcp78,silbey-jcp80,silbey-jcp83,harris-jcp83,nitzan,cheng-jcp128,jang-jcp129,jang-jcp131,jang-jcp135,nazir-prl103,mccutcheon-jcp35,zimanyi-ptrsa-370,yang-jcp137,pollock-njp15,nazir-jpcm28,pouthier-jcp138,chen-jpcb115,zhao-jcp137,chorosajev-jcp140,hamm-prb78,lee-jcp142,xu-fp11,wang-prb101,balzer-cs12}  has served as both an important conceptual framework and an efficient computational tool for describing charge and exciton transfer/transport dynamics in various condensed and complex media. PT creates a polaron picture where molecular vibrations and phonon modes of environments, collectively referred to as bath here, relax differently with respect to different site localized states.   When such responses to localized states are fast, the states ``dressed" with those responses, so called polarons, serve as effective means to represent  the contribution of the bath because they have already taken some contributions of the bath responses into consideration up to an infinite order.  Alternatively, PT can simply be viewed as a useful unitary transformation in the combined space of system and bath, which produces a new renormalized interaction Hamiltonian term that remains small even in the limit of strong couplings to the bath.  \re{As long as physical observables of interest remain unentangled with the bath through the PT, a quantum master equation (QME) derived by projecting out the bath in the polaron picture can be used for calculating such observables}.  Indeed, utilizing this fact has led to a general approach called polaron-transformed QME (PQME).\cite{jang-jcp129,jang-jcp131,jang-jcp135,nazir-prl103,mccutcheon-jcp35,zimanyi-ptrsa-370,yang-jcp137,pollock-njp15,nazir-jpcm28} 

The merit of PQME is its efficiency in accounting for both weak and strong couplings to the bath, while offering reasonable interpolation between the two limits.
Earlier versions of PQMEs\cite{jang-jcp129,jang-jcp131,jang-jcp135,nazir-prl103}  employed a full PT where all the bath degrees of freedom, modeled as harmonic oscillators, are fully relaxed to site local system states.    
However, in some cases, this is not \re{advantageous} because it invokes an over-relaxation of some slow modes, which can cause premature suppression of important coherence terms, \re{in particular, within the second order time-local approximation}. Variational PQME\cite{mccutcheon-jcp35,zimanyi-ptrsa-370,pollock-njp15,nazir-jpcm28} that combines a variational ansatz\cite{emery-prb9,silbey-jcp80,silbey-jcp83,harris-jcp83} with PQME and, more recently, frozen mode PQME\cite{teh-jcp150} approaches have been developed to address this issue.     This work develops a formulation that can put these works\cite{mccutcheon-jcp35,pollock-njp15,teh-jcp150} into a broader context and \re{can ameliorate} the issue of over-relaxation by extending the PQME for partial PTs of fairly general kind, where high frequency bath modes are transformed preferentially.  It is shown here that a general closed form expression of a 2nd order time-local PQME including all the inhomogeneous terms can be derived as in the case of the full PT,\cite{jang-jcp131} although the resulting expressions are more complicated.  All the terms needed for the calculation of a second order partially polaron-transformed QME (p-PQME) are derived in detail.  Then, numerical results are provided for a model with two system states, which demonstrate the feasibility and the utility of the p-PQME approach. 

The paper is organized as follows.  Section II provides the model Hamiltonian and presents derivation of all terms involved in the second order time-local p-PQME.  Section III considers a case of two-level system and presents  results of model calculations.  Section IV summarizes main results of this paper and offers concluding remarks.

\section{Theory}
\subsection{Hamiltonian and partial polaron transformation}
Let us consider a quantum system consisting of $N$ coupled quantum states, \re{$|j\rangle$'s}, with each representing a localized electronic excitation or charge carrying state.  
The Hamiltonian of the quantum system consisting of these states in general can be expressed as
\be
H_s=\sum_{j=1}^N E_j |j\rangle\langle j| + \sum_{j,k=1}^N  J_{jk} |j\rangle\langle k|,  \label{eq:h_s}
\ee
where $E_j$ is the energy of the site local state $|j\rangle$, and $J_{jk}$, for $j\neq k$, is the electronic coupling between states $|j\rangle$ and $|k\rangle$ that is assumed to be a real number here.  The double summation in the above equation also includes the case $j=k$, for which $J_{jk}$ is assumed to be zero.  

All other degrees of freedom constituting the total Hamiltonian are referred to here as bath, which includes all the vibrational modes of molecules and the polarization response of environmental degrees of freedom.    Some of these may exhibit significant anharmonic character in real molecular systems. However, for the sake of simplicity, it is assumed here that all of them can be modeled as coupled harmonic oscillators. In addition, all couplings of the bath to the  system Hamiltonian are assumed to be diagonal with respect to the site local states $|j\rangle$'s and to be linear in the displacements of the bath oscillators.  Thus, the  total Hamiltonian is assumed to be of the following standard form:
\be 
H=H_s+H_b+H_{sb} , \label{eq:hamil-tot}
\ee
where  
\ben
&&H_b=\sum_n \hbar\omega_n (b_n^\dagger b_n +\frac{1}{2}) , \label{eq:h_b}\\
&&H_{sb}=\sum_{j=1}^N\sum_n \hbar \omega_n g_{n,j}(b_n+b_n^\dagger ) |j\rangle\langle j| . \label{eq:h_sb} 
\een
In the above expressions, $\omega_n$ and $b_n$ ($b_n^\dagger$) are the frequency and the lowering (raising) operator of each normal mode constituting the bath degrees of freedom, and $g_{n,j}$ represents the (dimensionless) coupling strength of each mode to the state $|j\rangle$.  \re{The nature of this bath can be characterized collectively by introducing the following bath spectral density:
\be
{\mathcal J}_{jj'}(\omega)=\pi \hbar \sum_n \delta (\omega-\omega_n) \omega_n^2 g_{n,j}g_{n,j'} . \label{eq:bath-spd}
\ee}

The total Hamiltonian, Eq. (\ref{eq:hamil-tot}), is a straightforward generalization of the spin-boson Hamiltonian\cite{caldeira-ap149,weiss} and has been used widely for molecular excitons\cite{kenkre-reineker,may,jang-exciton} and charge transport dynamics.\cite{weiss,may,coropceanu-cr107}  The full information on the combined system and bath degrees of freedom in general requires determining the total density operator $\rho(t)$, which is governed by the following quantum Liouville equation:
\be
\frac{d}{dt} \rho(t)=-i {\mathcal L}\rho(t)\equiv -\frac{i}{\hbar}[H,\rho(t)] .
\ee
Due to the large number of bath degrees of freedom typically involved, solving the above equation exactly is difficult even for simple forms of $H_b$ and $H_{sb}$ given by Eqs. (\ref{eq:h_b}) and (\ref{eq:h_sb}).   In the quantum master equation approach, only the reduced system degrees of freedom is solved explicitly whereas the effects of the bath are treated only to certain extents that are necessary.  Typically, these effects can be encoded entirely into appropriate bath spectral densities.

For the case of Ohmic or sub-Ohmic bath spectral density, \re{namely, when Eq. (\ref{eq:bath-spd}) behaves linearly or sub-linearly in the low frequency limit,} the sum of Huang-Rhys factors for the bath degrees of freedom diverges.\cite{weiss}  This implies vanishing Debye-Waller factors, which result in premature suppression of  some coherence terms when the full PT is applied first.  This issue is significant in particular for charge transfer processes, where the bath spectral density is typically known to contain Ohmic or even sub-Ohmic low frequency components.  A simple way to avoid such suppression is to limit the PT to only fast enough bath modes.   To this end, let us introduce a weighting function $W_h(\omega)$ with the following limiting behavior:
\be
W_h(\omega) = \left \{ \begin{array}{cl} O(\omega^\alpha)  &\mbox{, for } \omega \rightarrow 0 \\
                                                         1 &\mbox{, for }  \omega \rightarrow \infty  \end{array} \right .      \label{eq:property-wh}
\ee  
The scaling behavior for $\omega \rightarrow 0$ in the above equation, for sufficiently large value of $\alpha$, can suppress the sluggish components of the bath spectral density and thus prevents the corresponding Debye-Waller factor from becoming zero.  For the case of Ohmic bath, \re{ this means that $\alpha \geq 1$ at least.}  If the bath spectral density has sub-Ohmic components, the lower bound for $\alpha$ should increase accordingly.  

Let us now define a generating function of a partial PT as follows:
\be
G=\sum_{j=1}^N\sum_n g_{n,j} W_h(\omega_n) (b_n^\dagger-b_n)|j\rangle\langle j| .
\ee
The corresponding PT, when applied to the system Hamiltonian, Eq. (\ref{eq:h_s}), results in
\ben
\tilde H_s&=&e^{G} H_s e^{-G} \nonumber \\
&=&\sum_{j=1}^N E_j|j\rangle\langle j|+\sum_{j,k=1}^N J_{jk} \theta_{j}^\dagger \theta_{k} |j\rangle\langle k| ,  \label{eq:tilde-hs}
\een
where 
\be
\theta_{j}=e^{-\sum_n g_{n,j} W_h(\omega_n) (b_n^\dagger-b_n)} .
\ee  
On the other hand, it is straightforward to show that 
\ben
&&e^{G} (H_b+H_{sb})e^{-G}  \nonumber \\
&&=H_b+\sum_{j=1}^N\sum_n  \hbar \omega_n g_{n,j}(1-W_h(\omega_n))(b_n+b_n^\dagger )|j\rangle\langle j| \nonumber \\
&&-\sum_{j=1}^N\sum_n \hbar\omega_n g_{n,j}^2 W_h(\omega_n)(2-W_h(\omega_n))|j\rangle\langle j|  .
\een
Combining the above expression with Eq. (\ref{eq:tilde-hs}), one can obtain the following expression for the transformed total Hamiltonian:
\be
\tilde H=e^{G} H e^{-G}=\tilde H_s^p+\tilde H_{sb} +H_b , \label{eq:tilde_h}
\ee 
where 
\be
\tilde H_s^p= \sum_{j=1}^N\tilde E_j |j\rangle\langle j|  . \label{eq:til_hs0}\\
\ee
\re{In the above expression, $\tilde E_j$ is a partially renormalized energy for site $j$ given by  
\be
\tilde E_j=E_j- \lambda_j , \label{eq:tilde-ej}
\ee
with $\lambda_j$ being the corresponding reorganization \re{energy} defined as
\ben
\lambda_j&=&  \sum_n \hbar \omega_n g_{n,j}^2 W_h(\omega_n)(2-W_h(\omega_n)) \nonumber \\
&=&\frac{1}{\pi} \int_0^\infty d\omega \frac{{\mathcal J}_{jj}(\omega)}{\omega} W_h(\omega)(2-W_h(\omega)) .
\een
In the second equality of the above equation, the definition of the bath spectral density, Eq. (\ref{eq:bath-spd}), was used.}

The second term in Eq. (\ref{eq:tilde_h}), $\tilde H_{sb}$, is  a partially renormalized system-bath interaction Hamiltonian given by
\ben 
&&\tilde H_{sb}=\sum_{j, k=1}^N  J_{jk} \theta_j^\dagger \theta_{k} |j\rangle\langle k|\  \nonumber \\
&&+\sum_{j=1}^N\sum_n \hbar\omega_n g_{n,j}(1-W_h(\omega_n))(b_n+b_n^\dagger)|j\rangle \langle j| ,\label{eq:til_hsb}
\een
where $\theta_{j}^\dagger \theta_{k} $ can be expressed as
\be
\theta_{j}^\dagger\theta_{k}=e^{\sum_n \delta g_{n,jk} W_h(\omega_n) (b_n^\dagger -b_n) }\ ,
\ee
with $\delta g_{n,jk}=(g_{n,j}-g_{n,k})$.

\subsection{Partially polaron transformed quantum master equation (p-PQME)}
A complete derivation of a QME defined in the partial polaron transformation  (p-PT) space, as defined in the previous subsection, is provided below. 

\subsubsection{Quantum Liouville equation in the polaron and interaction picture}
The total Hamiltonian in the partial polaron picture, Eq. (\ref{eq:tilde_h}),  can be divided into new effective zeroth and first order terms as follows: 
\be
\tilde H=\tilde H_0+\tilde H_1, \label{eq:tot-transform}
\ee
where
\ben
&&\tilde H_0=\tilde H_s^p+\langle \tilde H_{sb}\rangle_b +H_b \equiv \tilde H_{0,s}+H_b  ,  \label{eq:til_h0}\\
&&\tilde H_1=\tilde H_{sb}-\langle \tilde H_{sb}\rangle_b . \label{eq:til_h1}
\een
In above expressions, $\langle \tilde H_{sb}\rangle_b=Tr_b\{ \tilde H_{sb} \rho_b\}$ with $\rho_b=e^{-\beta H_b}/Tr_b\{ e^{-\beta H_b}\}$.  This term represents the average system-bath interaction in the partial polaron picture.  

In Eq. (\ref{eq:til_h0}), the system part of the new zeroth order Hamiltonian, $\tilde H_{0,s}=\tilde H^p_s+\langle \tilde H^c_s\rangle_b$, includes the average effect of system-bath interactions in the partial polaron picture, and can be expressed as
\be
\tilde H_{0,s}=\sum_{j=1}^N \tilde E_j |j\rangle\langle j|+\sum_{j,k=1}^N  \tilde J_{jk} |j\rangle\langle k| ,
\ee 
where $\tilde E_j$ has been defined by Eq. (\ref{eq:tilde-ej}) and $\tilde J_{jk}=w_{jk} J_{jk}$ with
\ben
w_{jk}&=&\langle \theta_{j}^\dagger \theta_{k}\rangle=\langle \theta_{k}^\dagger \theta_{j}\rangle\nonumber \\
&=&e^{-\sum_n \coth(\beta\hbar \omega_n/2)\delta g_{n,jk}^2W_h(\omega_n)^2/2} . \label{eq:w_jk}
\een
Unlike the case of the full PT, the Debye-Waller factor $w_{jk}$ given above is non-zero even for the Ohmic bath spectral density given that the weighting function satisfies Eq. (\ref{eq:property-wh}). 

Similarly, the first order term $\tilde H_1$, Eq. (\ref{eq:til_h1}), can be expressed as
\be
\tilde H_1=\sum_{j,k=1}^N \tilde B_{jk}|j\rangle \langle k|  , \label{eq:h1}
\ee
where \be
\tilde B_{jk}=J_{jk}(\theta_{j}^\dagger \theta_{k} -w_{jk})+\delta_{jk} D_{j} . \label{eq:b_def} 
\ee
In the above expression, $\delta_{jk}$ is the Kronecker-delta symbol and  
\be
D_{j}=\sum_n\hbar\omega_n g_{n,j}(1-W_h(\omega_n)) (b_n+b_n^\dagger) .  \label{eq:d_j}
\ee
Thus, the bath operator $\tilde B_{jk}$ given by Eq. (\ref{eq:b_def}) is a sum of the renormalized system-bath interaction term (relative to its average) due to partial PT (for $j\neq k$) and of the remaining linear interaction term (for $j=k$) for portions of bath modes that have not been transformed.

Having defined $\tilde H_0$, for which exact time evolution can be implemented numerically, let us now consider the dynamics in the interaction picture of $\tilde H_0$.  First, $\tilde H_1$ in this interaction picture becomes 
\ben
&&\tilde H_{1,I}(t)\equiv e^{i\tilde H_0 t/\hbar} \tilde  H_1 e^{-i\tilde H_0 t/\hbar}\nonumber \\
&&=\sum_{j, k=1}^N \tilde B_{jk} (t)  {\mathcal T}_{jk} (t) , \label{eq:h1I}
\een
where 
\ben
&&{\mathcal T}_{jk} (t)=e^{i\tilde H_{0,s} t/\hbar} |j\rangle \langle k| e^{-i\tilde H_{0,s} t/\hbar} ,\label{eq:t_jk} \\
&&\tilde B_{jk}(t)=e^{iH_bt/\hbar} \tilde B_{jk} e^{-iH_bt/\hbar} \nonumber \\
&&=J_{jk}(\theta_{j}^\dagger (t)\theta_{k}(t)-w_{jk})+\delta_{jk} D_{j}(t)   . \label{eq:b_jk-t}
\een
In the above expression, 
\ben
&&\theta_{j}^\dagger (t)\theta_{k}(t)= e^{\sum_n \delta g_{n,jk} W_h(\omega_n)(b_n^\dagger e^{i\omega_nt} - b_ne^{-i\omega_n t})} , \label{eq:bl_t} \\
&&D_{j}(t)= \sum_n\hbar\omega_n g_{n,j} (1-W_h(\omega_n))\nonumber \\
&&\hspace{1in} \times (b_n e^{-i\omega_n t}+b_n^\dagger e^{i\omega_n t}) .
\een

In the interaction picture with respect to $\tilde H_0$, the partially polaron-transformed total density operator becomes $\tilde \rho_I (t)=e^{i\tilde {\mathcal L}_0t} \tilde \rho (t)$, which evolves according to the following time dependent quantum Liouville equation:
\be
\frac{d}{dt}\tilde \rho_I(t) =-i\tilde {\mathcal L}_{1,I}(t)\tilde \rho_I (t)\equiv  -\frac{i}{\hbar} [\tilde H_{1,I} (t),\tilde \rho_I (t)] \ , \label{eq:rho_i_lv}
\ee
where the second equality serves as the definition of $\tilde {\mathcal L}_{1,I}(t)$.

\subsubsection{Quantum master equation for reduced density operator}
Taking trace of $\tilde \rho_I(t)$ over the bath degrees of freedom leads to the following interaction-picture reduced system density operator defined in the p-PT system-bath space:
\be
\tilde \sigma_I(t)\equiv Tr_b\left\{ \tilde \rho_I (t)\right\} .
\ee
While the above reduced density operator still retains full information on the  system degrees of freedom in the p-PT space, it is important to note that the trace operation makes it impossible to retrieve the full information on the system prior to the application of p-PT.  On the other hand, properties diagonal in the site basis, which are not affected by p-PT,  remain intact.   

A formally exact time evolution equation with time-convolution can be obtained for $\tilde \sigma_I(t)$ employing the standard projection operator technique\cite{jang-jcp116,jang-exciton} for a well-known projection operator ${\mathcal P}(\cdot)\equiv \rho_b Tr_b\{(\cdot)\}$ as follows:
\ben
&&\frac{d}{dt}\tilde \sigma_I(t)=-\int_0^t d\tau Tr_b\Big \{ \tilde {\mathcal L}_{1,I}(t) e_{(+)}^{-i\int_\tau^t d\tau' {\mathcal Q} \tilde {\mathcal L}_{1,I}(\tau')}  \nonumber \\ && \hspace{1.5in}\times  {\mathcal Q} \tilde {\mathcal L}_{1,I}(\tau) \rho_b\Big \}\tilde \sigma_I(\tau) \nonumber \\
&&\hspace{.5in}-iTr_b\{ \tilde {\mathcal L}_{1,I}(t)e_{(+)}^{-i\int_0^t d\tau {\mathcal Q} \tilde {\mathcal L}_{1,I}(\tau)} {\mathcal Q} \tilde \rho(0) \} , \label{eq:exact-tn}
\een
where \re{$e_{(+)}$ denotes the exponential operator with chronological time ordering}, ${\mathcal Q}=1-{\mathcal P}$ and 
\be
{\mathcal Q}\tilde \rho (0)=e^G \rho(0)e^{-G}-\rho_b Tr_b\{e^G\rho(0)e^{-G}\}  . \label{eq:qrho_0}
\ee
Alternatively, replacing $\tilde \rho_I(\tau)$ with the back propagation of $\tilde \rho_I(t)$ from $t$ to $\tau$ within the projection operator formalism,\cite{shibata-jpsj49,jang-exciton} one can obtain the following formally exact time-local equation:
\ben
&&\frac{d}{dt}\tilde \sigma_I(t)=-\int_0^t d\tau Tr_b\Big \{ \tilde {\mathcal L}_{1,I}(t) (1+i\Gamma_{1,I}(t))^{-1}  \nonumber \\ && \hspace{.2in}\times  e_{(+)}^{-i\int_\tau^t d\tau' {\mathcal Q} \tilde {\mathcal L}_{1,I}(\tau')} {\mathcal Q} \tilde {\mathcal L}_{1,I}(\tau) {\mathcal P}e_{(-)}^{i\int_{\tau}^t d\tau' \tilde {\mathcal L}_{1,I}(\tau')} \rho_b\Big \}\tilde \sigma_I(t) \nonumber \\
&&\hspace{.1in}-i{\mathcal P} \tilde {\mathcal L}_{1,I}(t)  (1+i\Gamma_{1,I}(t))^{-1} e_{(+)}^{-i\int_0^t d\tau {\mathcal Q} \tilde {\mathcal L}_{1,I}(\tau)} {\mathcal Q} \tilde \rho(0)  ,  \label{eq:qme-2nd}
\een
where \re{$e_{(-)}$ denotes the exponential operator with anti-chronological time ordering and}
\be
\Gamma_{1,I}(t)=\int_0^t d\tau e_{(+)}^{-i\int_{\tau}^t d\tau' {\mathcal Q} \tilde {\mathcal L}_{1,I}(\tau')}{\mathcal Q} \tilde {\mathcal L}_{1,I}(\tau){\mathcal P} e_{(-)}^{i\int_\tau^t d\tau' \tilde {\mathcal L}_{1,I}(\tau')} .
\ee
When approximated up to the second order, it is straightforward to show that Eq. (\ref{eq:qme-2nd}) simplifies to 
\be
\frac{d}{dt} \tilde \sigma_I (t) =- {\mathcal R} (t)\tilde \sigma_I(t) +{\mathcal I}(t) , \label{eq:qme_tl-2nd}
\ee
where 
\ben
{\mathcal R}(t)&=&\int_0^t d\tau Tr_b\{ \tilde{ \mathcal L}_{1,I}(t) \tilde{ \mathcal L}_{1,I}(\tau) \rho_b\}  ,  \label{eq:rt_def}\\
{\mathcal I}(t)&=&{\mathcal I}^{(1)}(t)+{\mathcal I}^{(2)}(t)\nonumber  \\
&=&-i Tr_b \{ \tilde {\mathcal L}_{1,I} (t) {\mathcal Q} \tilde \rho (0)\} \nonumber \\
&&-\int_0^t d\tau Tr_b \{\tilde {\mathcal L}_{1,I} (t) \tilde {\mathcal L}_{1,I}(\tau) {\mathcal Q} \tilde \rho (0)\} . \label{eq:it}
\een
In the above expression, ${\mathcal I}^{(1)}(t)$ and ${\mathcal I}^{(2)}(t)$ represent the first and second order inhomogeneous terms of the time evolution equation.   Note that Eq. (\ref{eq:qme_tl-2nd}) can also be obtained from  Eq. (\ref{eq:exact-tn}), by simply replacing $\tilde \sigma_I(\tau)$ with $\tilde \sigma_I(t)$, which does not affect the accuracy at the second order level.

Equation (\ref{eq:qme_tl-2nd}) is the 2nd order time local p-PQME expressed in Liouville space.   In all previous works and in the present paper, this time local form has been chosen due to its convenience.  However, a time nonlocal 2nd order expressions can also be derived directly from Eq. (\ref{eq:exact-tn}), and its performance compared to the time-local form needs to be understood better through actual numerical studies.   Many numerical tests so far seem to indicate that the performance of the time-local form is better than that of the time non-local form for exciton and charge transfer dynamics near room temperature.  However, \re{considering that the 2nd order time-nonlocal PQME is equivalent to  the non-interacting blip approximation,\cite{leggett-rmp59,silbey-jcp80,aslangul-pl110a,dekker-pra35} which accounts for significant contribution of coherent dynamics even for the Ohmic bath, it is likely that the time non-local equation becomes more reliable as the bath becomes sluggish.}   In addition, a recent work\cite{lai-jcp155} also provides examples of the case where the performance of time-nonlocal QME is more satisfactory than the time-local form.  Therefore, further tests and comparative calculations are necessary to make more comprehensive assessment of the two approaches.  With this point clear, the rest of this section provides detailed expressions for the relaxation superoperator,  Eq. (\ref{eq:rt_def}), and the inhomogeneous term, Eq. (\ref{eq:it}), in the Hilbert space.

\subsubsection{Homogeneous terms of the  2nd order time local p-PQME}
The Hilbert space expression for ${\mathcal R} (t)\tilde \sigma_I(t)$ in Eq. (\ref{eq:qme_tl-2nd}) can be obtained by employing Eqs. (\ref{eq:h1I})-(\ref{eq:b_jk-t}) in Eq. (\ref{eq:rho_i_lv}) and by taking advantage of the cyclic invariance of the trace operation with respect to the bath degrees of freedom.  The resulting expression is as follows:
\ben
{\mathcal R}(t) \tilde \sigma_I (t)&=&\frac{1}{\hbar^2}\sum_{j, k=1}^N\sum_{j', k'=1}^N \int_0^t d\tau\langle \tilde B_{jk} (t)\tilde B_{j'k'} (\tau)\rangle \nonumber \\
&&\hspace{0.8in} \times  [{\mathcal T}_{jk} (t), {\mathcal T}_{j'k'} (\tau) \tilde \sigma_I (t)] \nonumber \\ 
&&+{\rm H.c.}  ,  \label{eq:qme_hom}
\een
where ${\rm H.c}$ refers to Hermitian conjugates of all previous terms and $\langle \tilde B_{jk}(t)\tilde B_{j'k'}(\tau)\rangle$ (with subscript $b$ omitted) represents averaging over the equilibrium bath density operator, $\rho_b$.   

Appendix A describes calculation of all the terms constituting $\langle \tilde B_{jk}(t)\tilde B_{j'k'}(\tau)\rangle$. When the resulting expressions, Eqs. (\ref{eq:theta-4}), (\ref{eq:djt-1}), (\ref{eq:b3-1}), and (\ref{eq:dd-t}), are used in Eq. (\ref{eq:bb-cor}), it can be expressed as 
\ben
 &&\langle \tilde B_{jk}(t)\tilde B_{j'k'}(\tau)\rangle=\tilde J_{jk}\tilde J_{j'k'} \left (e^{-{\mathcal K}_{jk,j'k'}(t-\tau)}-1\right)  \nonumber \\
 &&\hspace{.5in}+\delta_{jk}\tilde J_{j'k'} {\mathcal M}_{j,j'k'}(t-\tau) +\delta_{j'k'}\tilde J_{jk} {\mathcal M}_{j',kj}(t-\tau) \nonumber \\
 &&\hspace{.5in}+\delta_{jk}\delta_{j'k'} {\mathcal C}_{jj'}(t-\tau) ,
  \label{eq:blm_eq}
 \een
 where
 \ben
 &&{\mathcal K}_{jk,j'k'}(t)= \sum_n \delta g_{n,jk} \delta g_{n,j'k'} W_h(\omega_n)^2 \nonumber\\ 
 &&\hspace{.3in} \times \left (\coth \left (\frac{\beta \hbar\omega_n}{2} \right ) \cos (\omega_n t) -i \sin (\omega_n t)\right) ,\label{eq:kt} \\
&& {\mathcal M}_{j,j'k'}(t)=\sum_n \hbar\omega_n g_{n,j}\delta g_{n,j'k'} (1-W_h(\omega_n)) W_h(\omega_n)\nonumber \\
 &&\hspace{.3in}\times \left (\cos(\omega_n t)-i\coth \left (\frac{\beta\hbar\omega_n}{2}\right ) \sin (\omega_n t) \right) , \label{eq:mt}\\
 && {\mathcal C}_{jj'}(t)=\sum_n \hbar^2\omega_n^2 g_{n,j}g_{n,j'} (1-W_h(\omega_n))^2 \nonumber \\
 &&\hspace{.3in}\times \left (\coth \left (\frac{\beta\hbar\omega_n}{2}\right )\cos(\omega_n t)-i \sin (\omega_n t) \right) . \label{eq:ct}
 \een
 Similar expressions as above have also been derived in the context of variational PQME.\cite{pollock-njp15}
Note that the three bath correlation functions defined above satisfy the following symmetry properties: 
 \ben
 {\mathcal K}_{jk,j'k'}(t)&=&{\mathcal K}_{kj,k'j'}(t)\nonumber \\
 &=&-{\mathcal K}_{jk,k'j'}(t)=-{\mathcal K}_{kj,j'k'}(t) ,   \\
 {\mathcal M}_{j,j'k'}(t)&=&-{\mathcal M}_{j,k'j'}(t) ,\\ 
 {\mathcal C}_{jj'}(t)&=&{\mathcal C}_{j'j}(t) . 
\een

\re{The three correlation functions, Eqs. (\ref{eq:kt})-(\ref{eq:ct}), can all be expressed in terms of the bath spectral density, Eq.(\ref{eq:bath-spd}).  For more compact expressions of these, let us introduce }the following auxiliary bath spectral densities:
\ben
&&{\mathcal J}^{(1)}_{j,j'k'}(\omega)=\pi\hbar \sum_n \delta (\omega-\omega_n) \omega_n^2 g_{n,j}\delta g_{n,j'k'} \nonumber \\
&&={\mathcal J}_{jj'}(\omega)-{\mathcal J}_{jk'}(\omega) ,\\
&&{\mathcal J}^{(2)}_{jk,j'k'}(\omega)=\pi\hbar \sum_n \delta (\omega-\omega_n) \omega_n^2 \delta g_{n,jk}\delta g_{n,j'k'} \nonumber \\
&&={\mathcal J}_{jj'}(\omega)+{\mathcal J}_{kk'}(\omega)-{\mathcal J}_{jk'}(\omega)-{\mathcal J}_{kj'}(\omega)  .
\een
Then, it is straightforward to show that Eqs. (\ref{eq:kt})-(\ref{eq:ct}) can be expressed as 
\ben
&&{\mathcal K}_{jk,j'k'}(t)=\frac{1}{\pi\hbar}\int_0^\infty d\omega \frac{{\mathcal J}^{(2)}_{jk,j'k'}(\omega)}{\omega^2} W_h(\omega)^2 \nonumber \\
&&\hspace{.5in}\times \left (\coth\left (\frac{\beta\hbar\omega}{2}\right)\cos (\omega t)-i\sin (\omega t)\right ) , \\
&&{\mathcal M}_{j,j'k'}(t)=\frac{1}{\pi} \int_0^\infty d\omega \frac{{\mathcal J}^{(1)}_{j,j'k'}(\omega)}{\omega}  (1-W_h(\omega))W_h(\omega) \nonumber \\
&&\hspace{.5in} \times \left (\cos (\omega t)-i \coth\left (\frac{\beta\hbar\omega}{2}\right)\sin (\omega t)\right ) , \\
&&{\mathcal C}_{jj'}(t)=\frac{\hbar}{\pi}\int_0^\infty d\omega {\mathcal J}_{jj'}(\omega) (1-W_h(\omega))^2 \nonumber \\
&&\hspace{.5in}\times \left (\coth\left (\frac{\beta\hbar\omega}{2}\right)\cos (\omega t)-i\sin (\omega t)\right ) . 
\een
Note also that $w_{jk}$ defined by Eq. (\ref{eq:w_jk}) can be expressed as follows:
\ben
&&w_{jk}=e^{-{\mathcal K}_{jk,jk}(0)/2}\nonumber \\
&&=\exp\left\{-\frac{1}{2\pi\hbar}\int_0^\infty d\omega \frac{{\mathcal J}^{(2)}_{jk,jk}(\omega)}{\omega^2} W_h(\omega)^2 \coth\left (\frac{\beta\hbar\omega}{2}\right)\right\} \nonumber \\
\een

If $W_h(\omega)=1$, ${\mathcal M}_{j,j'k'}(t)$ and ${\mathcal C}_{jj'}(t)$ become zero and the expressions for ${\mathcal K}_{jk,j'k'}(t)$ and $w_{jk}$ reduce to those for the original 2nd order PQME based on the full PT.    On the other hand, for $W_h(\omega)=0$, ${\mathcal K}_{jk,j'k'}(t)$ and ${\mathcal C}_{jj'}(t)$ become zero and the expression for ${\mathcal C}_{jj'}(t)$ reduce to that for a conventional 2nd order time-local QME (without PT). In this sense, the above expressions can be viewed as general ones incorporating the two limiting cases.  As the next simplest case, let us consider the case where $W_h(\omega)$ is a step function. For this case, $W_h(\omega)(1-W_h(\omega))=0$ and thus ${\mathcal M}_{j,j'k'}(t)$ becomes zero.  As a result, the relaxation superoperator becomes  a simple sum of those due to p-PT part and untransformed linear coupling terms.

\subsubsection{Inhomogeneous terms of p-PQME}
For the calculation of inhomogeneous terms, let  us assume that the  initial untransformed total density operator is given by 
\be
\rho(0)=\sigma(0)\rho_b ,
\ee
where
\be
\sigma(0)=\sum_{j,k=1}^N\sigma_{jk}(0)|j\rangle\langle k| .
\ee
Then, 
\be 
{\mathcal Q}\tilde \rho(0) =\sum_{j,k=1}^N \sigma_{jk}(0) {\mathcal T}_{jk}(0) \delta \tilde \rho_{b,jk}  ,
\ee
where Eq. (\ref{eq:t_jk}) with $t=0$ has been used and 
\be
\delta \tilde \rho_{b,jk}=\theta^\dagger_j \rho_b \theta_k -w_{jk} \rho_b  , \label{eq:delta-rho}
\ee
with the convention that $w_{jk}=1$ for $j=k$.
Therefore, the first order inhomogeneous term in Eq. (\ref{eq:it}) can be expressed as
\ben
&&{\mathcal I}^{(1)}(t)=-iTr_b\{ \tilde {\mathcal L}_{1,I}(t) {\mathcal Q} \tilde \rho (0)\}\nonumber \\
&&\hspace{.5 in} =-\frac{i}{\hbar}\sum_{j,k=1}^N\sum_{j',k'=1}^N Tr_b\left\{\tilde B_{jk}(t)\delta \tilde \rho_{b,j'k'} \right\} \nonumber \\
&&\hspace{1.2in} \times \sigma_{j'k'}(0) [{\mathcal T}_{jk}(t),{\mathcal T}_{j'k'}(0)] . \label{eq:ihom-1}
\een
In the above expression, the trace over the bath can be calculated explicitly employing the definitions for $\tilde B_{jk}(t)$ and $\delta \tilde \rho_{b,jk}$, Eqs. (\ref{eq:b_jk-t}) and (\ref{eq:delta-rho}), respectively.   Details of this calculation are provided in Appendix B, and the resulting expression can be summed up as
\ben
&&Tr_b\{ \tilde B_{jk} (t)\delta \tilde \rho_{b,j'k'} \} \nonumber \\
&&=w_{j'k'} \Big \{ \tilde J_{jk} \left ( e^{-{\mathcal K}_{jk,j'k'}(t)} f_{jk,k'} (t)-1 \right) \nonumber \\
&&\hspace{.5in}+\delta_{jk}\left ( {\mathcal M}_{j,j'k'}(t)+ h_{j,k'}(t)\right ) \Big \} , \label{eq:ihom1-bath}
\een
where 
\ben
&&f_{jk,k'}(t)=\exp\left \{2i\sum_n g_{n,k'}\delta g_{n,jk} W_h(\omega_n)^2 \sin (\omega_n t)\right\}\nonumber \\
&&=\exp\left \{\frac{2i}{\pi\hbar} \int_0^\infty d\omega \frac{{\mathcal J}^{(1)}_{k',jk} (\omega)}{\omega^2} W_h(\omega)^2 \sin (\omega t) \right\} ,  \label{eq:ft} \\
&&h_{j,k'}(t)=2\sum_n \hbar\omega_n g_{n,j}g_{n,k'} (1-W_h(\omega_n))W_h(\omega_n) \cos(\omega_n t) \nonumber \\
&&=\frac{2}{\pi} \int_0^\infty d\omega \frac{{\mathcal J}_{jk'}(\omega)}{\omega} (1-W_h(\omega))W_h(\omega) \cos (\omega t) .  \label{eq:ht}
\een

Similarly, the second order inhomogeneous term in Eq. (\ref{eq:it}) can be expressed as
\ben
{\mathcal I}^{(2)}(t)&=&-\int_0^t d\tau Tr_b\left \{ \tilde {\mathcal L}_{1,I}(t)\tilde {\mathcal L}_{1,I}(\tau) {\mathcal Q} \tilde \rho(0) \right\} \nonumber \\
&=&-\frac{1}{\hbar^2} \sum_{j,k=1}^N\sum_{j',k'=1}^N\sum_{j'',k''=1}^N \nonumber \\
&& \int_0^t d\tau Tr_b \left \{\tilde B_{jk}(t)\tilde B_{j'k'}(\tau) \delta \tilde \rho_{b,j''k''} \right\} \nonumber \\
&&\times \sigma_{j''k''}(0) [{\mathcal T}_{jk}(t), {\mathcal T}_{j'k'}(\tau) {\mathcal T}_{j''k''}(0)]  + {\rm H.c.} , \nonumber \\ \label{eq:ihom2}
\een
where H.c. refers to Hermitian conjugates of all previous terms.
Detailed expressions for the trace of bath operators in the above expression are derived in Appendix C. 

As in the case of the relaxation superoperator, the expressions for the inhomogeneous terms shown above reduce to those for the 2nd order PQME for $W_h(\omega)=1$ and those for the regular 2nd order QME for $W_h(\omega)=0$.  For the case where $W_h(\omega)(1-W_h(\omega))=0$, they become independent sums of those due to PT and due to untransformed  linear system-bath couplings. 

\subsection{Representation in the basis of renormalized system eigenstates}
For both better conceptual understanding and efficient numerical calculation, it is convenient to consider the dynamics in the basis of eigenstates of $\tilde H_{0,s}$.  Let us  denote the $p$th eigenstate and eigenvalue of $\tilde H_{0,s}$ as $|\varphi_p\rangle$ and ${\mathcal E}_p$. Then, 
\be
\tilde H_{0,s}=\sum_{p=1}^N {\mathcal E}_p |\varphi_p\rangle\langle \varphi_p| \ .
\ee
The transformation matrix $U$ between the site localized states $|j\rangle$'s and the eigenstates $|\varphi_p\rangle$'s can be defined such that $U_{jp}=\langle j|\varphi_p\rangle$. Then, 
\be
|j\rangle=\sum_{p=1}^N U_{jp}^* |\varphi_p\rangle \ . \label{eq:lj_trans}
\ee
This transformation can be used to express ${\mathcal T}_{jk}(t)$ defined by Eq. (\ref{eq:t_jk}) in the basis of $|\varphi_p\rangle$'s as follows:  
\be
{\mathcal T}_{jk}(t)=\sum_{p,q=1}^N U_{jp}^* U_{kq} e^{i\delta {\mathcal E}_{pq} t/\hbar} |\varphi_p\rangle \langle \varphi_{q}| \ , \label{eq:tl_ex}
\ee 
where $\delta {\mathcal E}_{pq}={\mathcal E}_p-{\mathcal  E}_{q}$.  Let us also introduce $S_{pq}(t)$'s such that 
\be
\tilde \sigma_I (t) =\sum_{p,q=1}^N S_{pq} (t) |\varphi_p\rangle \langle \varphi_{q}| . \label{eq:sigm_ex}
\ee
Then, with some arrangement of dummy summation indices, it is straightforward to show that 
\ben
&&[{\mathcal T}_{jk}(t),{\mathcal T}_{j'k'}(\tau)\tilde \sigma_I(t)] \nonumber \\
&&=\sum_{p,q=1}^N |\varphi_p\rangle\langle \varphi_q| \sum_{p',q'=1}^N \Big (\delta_{q'q} \sum_{r=1}^N  U_{jp}^*U_{kr}U_{j'r}U_{k'p'}   \nonumber\\
&&\hspace{1.5in}\times e^{i\delta {\mathcal E}_{pp'} t/\hbar} e^{i\delta {\mathcal E}_{p'r} (t-\tau)/\hbar} \nonumber \\
&&\hspace{1in}-U_{jq'}^*U_{kq}U_{j'p}^*U_{k'p'} e^{i(\delta {\mathcal E}_{pp'}-\delta {\mathcal E}_{qq'})t/\hbar}\nonumber \\
&&\hspace{1.5in}\times e^{i\delta {\mathcal E}_{p'p} (t-\tau)/\hbar} \Big ) S_{p'q'}(t)  . \label{eq:tt-sig-exp}
\een
The above expression, in combination with Eq. (\ref{eq:blm_eq}), can be used to express ${\mathcal R}(t)\tilde \sigma_I(t)$ in the basis of $|\varphi_p\rangle$'s. 
For more compact expression, let us introduce 
\ben
&&{\mathcal W}_{jk,j'k'}^{pq}(t)=\tilde J_{jk}\tilde J_{j'k'} \int_0^t d\tau\  e^{i\delta {\mathcal E}_{pq}(t-\tau)/\hbar} \nonumber \\ 
&&\hspace{1.2in}\times \Big ( e^{-{\mathcal K}_{jk,j'k'}(t-\tau)}-1\Big ) ,  \label{eq:W-def}\\
&&{\mathcal Y}_{j,j'k'}^{pq} (t) =\tilde J_{j'k'} \int_0^t d\tau e^{i\delta {\mathcal E}_{pq} (t-\tau)/\hbar} {\mathcal M}_{j,j'k'} (t-\tau) , \label{eq:Y-def}\\
&&{\mathcal X}_{jj'}^{pq}(t)=\int_0^t d\tau e^{i\delta {\mathcal E}_{pq}(t-\tau)/\hbar} {\mathcal C}_{jj'} (t-\tau) . \label{eq:X-def}
\een
Then,  
\be
{\mathcal R}(t)\tilde \sigma_I (t) =\sum_{p,q=1}^N\sum_{p',q'=1}^N |\varphi_p\rangle\langle \varphi_q| {\mathcal R}_{pq}^{p'q'} (t)S_{p'q'}(t) ,
\ee 
where
\ben 
&&{\mathcal R}^{p'q'}_{pq}(t)=\frac{1}{\hbar^2}\sum_{j,k=1}^N\sum_{j',k'=1}^N\nonumber \\
&& \hspace{.5in}\times \Bigg \{\delta_{q'q} \sum_{r=1}^N U_{jp}^*U_{kr}U_{j'r}^*U_{k'p'} e^{i\delta {\mathcal E}_{pp'} t/\hbar}  \nonumber \\
&&\hspace{1 in} \times \left ({\mathcal W}_{jk,j'k'}^{p'r}(t)+\delta_{jk} {\mathcal Y}_{j,j'k'}^{p'r}(t) \right .\nonumber \\
&&\hspace{1 in}\left . +\delta_{j'k'} {\mathcal Y}_{j',kj}^{p'r}(t)+\delta_{jk}\delta_{j'k'} {\mathcal X}_{jj'}^{p'r}(t) \right ) \nonumber \\
&&\hspace{.7in} - U_{jq'}^*U_{kq}U_{j'p}^*U_{k'p'} e^{i(\delta {\mathcal E}_{pp'}-\delta {\mathcal E}_{qq'})t/\hbar}\nonumber \\
&&\hspace{1 in} \times \left ({\mathcal W}_{jk,j'k'}^{p'p}(t)+\delta_{jk} {\mathcal Y}_{j,j'k'}^{p'p}(t) \right .\nonumber \\
&&\hspace{1 in} \left . +\delta_{j'k'} {\mathcal Y}_{j',kj}^{p'p}(t)+\delta_{jk}\delta_{j'k'} {\mathcal X}_{jj'}^{p'p}(t) \right ) \Bigg \}\nonumber \\
&&\hspace{.5in}+[{\rm c.c.}, p\leftrightarrow q, p'\leftrightarrow q']  . \label{eq:r-pq}
\een 
In the above expression, the last line represents complex conjugates along with the interchange of indices, $p \leftrightarrow q$ and $p' \leftrightarrow q'$, of all previous terms. 

For the calculation of the first order inhomogeneous term, the commutator of system operators in Eq. (\ref{eq:ihom-1}) can be calculated in a manner similar to Eq. (\ref{eq:tt-sig-exp}).  The resulting expression is 
\ben
[{\mathcal T}_{jk}(t),{\mathcal T}_{j'k'}(0)]&=&\sum_{p,q=1}^N|{\varphi}_q\rangle\langle \varphi_q| \nonumber \\
&&\times \sum_r \left ( U_{jp}^*U_{kr}U_{j'r}^*U_{k'q} e^{i\delta {\mathcal E}_{pr} t/\hbar } \right .\nonumber  \\
&&\left . -U_{jr}^*U_{kq}U_{j'p}^*U_{k'r} e^{i\delta {\mathcal E}_{rq} t/\hbar } \right ) ,
\een
which can also be obtained from Eq. (\ref{eq:tt-sig-exp}) by replacing $S_{p'q'}(t)$ with $\delta_{p'q'}$ and assuming $\tau=0$.
Combining the above expression with Eq. (\ref{eq:ihom1-bath}), one can express the first order inhomogeneous term, Eq. (\ref{eq:ihom-1}), as follows:
\ben
{\mathcal I}^{(1)}(t)&=&-\frac{i}{\hbar}\sum_{p,q=1}^N |\varphi_p\rangle \langle \varphi_q| \sum_{r=1}^N\sum_{j,k=1}^N\sum_{j',k'=1}^N \sigma_{j'k'}(0) w_{j'k'}\nonumber \\
 &&\times\left (U_{jp}^*U_{kr}U_{j'r}^* U_{k'q} e^{i\delta {\mathcal E}_{pr}t/\hbar} \right . \nonumber \\
 &&\hspace{.2in} \left . -U_{jr}^*U_{kq}U_{j'p}^*U_{k'r} e^{i\delta {\mathcal E}_{rq} t/\hbar} \right ) \nonumber \\
 &&\times \Big \{\tilde J_{jk} (e^{-{\mathcal K}_{jk,j'k'}(t)}f_{jk,k'}(t)-1) \nonumber \\
 &&\hspace{.2in}+\delta_{jk}\left (M_{j,j'k'}(t)+h_{j,k'}(t) \right )\Big\} . \label{eq:i1-exciton}
 \een

Similar expressions for the second order inhomogeneous term, Eq. (\ref{eq:ihom2}), can be obtained by replacing $S_{p'q'}(t)$ in Eq. (\ref{eq:tt-sig-exp}) with $U_{j''p'}^*U_{k''q'}$.   The resulting expression is as follows:
\ben
&&[{\mathcal T}_{jk}(t),{\mathcal T}_{j'k'}(\tau){\mathcal T}_{j''k''}(0)] \nonumber \\
&&=\sum_{p,q=1}^N |\varphi_p\rangle\langle \varphi_q| \sum_{p',q'=1}^N \Big (\delta_{q'q} \sum_{r=1}^N  U_{jp}^*U_{kr}U_{j'r}U_{k'p'}   \nonumber\\
&&\hspace{1.in}\times e^{i\delta {\mathcal E}_{pp'} t/\hbar} e^{i\delta {\mathcal E}_{p'r} (t-\tau)/\hbar} \nonumber \\
&&\hspace{0.5in}-U_{jq'}^*U_{kq}U_{j'p}^*U_{k'p'} e^{i(\delta {\mathcal E}_{pp'}-\delta {\mathcal E}_{qq'})t/\hbar}\nonumber \\
&&\hspace{1.in}\times e^{i\delta {\mathcal E}_{p'p} (t-\tau)/\hbar} \Big ) U_{j'' p'}^* U_{k''q'}  .  \label{eq:inhom2-eigen}
\een
Combining the above expression with bath correlation functions introduced in Appendix C, one can express the second order inhomogeneous term as follows:
\ben
{\mathcal I}^{(2)}(t)&=&-\sum_{j,k=1}^N\sum_{j',k'=1}^N\sum_{j'',k''=1}^N\sigma_{j''k''}(0) \sum_{p,q=1}^N|\varphi_p\rangle\langle \varphi_q| \nonumber \\
&&\times \sum_{p',q'=1}^N \sum_{r=1}^NU_{j''p'}^*U_{k''q'} e^{i\delta {\mathcal E}_{pp'} t/\hbar}  \nonumber \\
&&\times \Big ( \delta_{q'q} U_{jp}^*U_{kr}U_{j'r}U_{k'p'}   \nonumber \\
&&\hspace{.2in}- \delta_{rp}U_{jq'}^*U_{kq}U_{j'p}^*U_{k'p} e^{-i\delta {\mathcal E}_{qq'}t/\hbar} \Big )\nonumber \\
&&\hspace{.2in}\times \Big (J_{jk}J_{j'k'} \tilde F_{jk,j'k'}^{j''k''}(t;\delta {\mathcal E}_{p'r}) \nonumber \\
&&\hspace{.4in}+\delta_{jk}J_{j'k'} \tilde H_{j,j'k'}^{(1),j''k''}(t;\delta {\mathcal E}_{p'r}) \nonumber \\
&&\hspace{.4in}+J_{jk}\delta_{j'k'} \tilde H_{jk,j'}^{(2),j''k''}(t;\delta {\mathcal E}_{p'r}) \nonumber \\
&&\hspace{.4in}+\delta_{jk}\delta_{j'k'} \tilde L_{j,j'}^{j''k''}(t;\delta {\mathcal E}_{p'r}) \Big )\nonumber \\
&&+{\rm H.c.} , \label{eq:i2-exciton}
\een 
where the definitions of Eqs. (\ref{eq:til-f1})-(\ref{eq:til-f4}) in Appendix C have been used.

\section{Two-level system with independent bath}
\subsection{Model and general expressions}
Let us consider the simplest case where there are two site local system states ($N=2$), with only one electronic coupling $J=J_{12}=J_{21}$ and only one Debye-Waller factor, $w=w_{12}=w_{21}$.    Then, introducing $\theta=\theta_1^\dagger\theta_2$ for the present case,  the renormalized zeroth order system Hamiltonian and the first order term of the Hamiltonian defined by Eq. (\ref{eq:h1}) reduce to
\ben
\tilde H_{0,s}&=& \tilde E_1|1\rangle\langle 1|+\tilde E_2|2\rangle\langle 2| +Jw (|1\rangle\langle 2|+|2\rangle\langle 1|) ,\\
\tilde H_1&=&D_1|1\rangle\langle 1|+D_2|2\rangle\langle 2| \nonumber \\
&+&J(\theta-w)|1\rangle\langle 2|+J(\theta^\dagger-w)|2\rangle\langle 1| ,
\een
where $w=\langle\theta\rangle=\langle \theta^\dagger\rangle$, and $D_1$ and $D_2$ have been defined by Eq. (\ref{eq:d_j}). 
Note also that, for the present case,
\be
\tilde J_{12}=\tilde J_{21}=wJ ,  
\ee 
and $\tilde J_{11}=\tilde J_{22}=0$ by definition.  The eigenvalues of $\tilde H_{0,s}$ are given by 
\re{\be
{\mathcal E}_{1,2}=\frac{\tilde E_1+ \tilde E_2}{2} \pm \frac{\tilde E_1-\tilde E_2}{2} \sec (2\xi)  , \\
\ee
with subscripts $1$ and $2$ on the lefthand side denoting respectively the $+$ and $-$ signs on the right hand side, and $\xi =\tan^{-1}(2Jw/(\tilde E_1-\tilde E_2))/2$.  
The corresponding eigenstates are given by 
\ben
&&|\varphi_1\rangle =\cos \xi |1\rangle +\sin \xi |2\rangle , \\
&&|\varphi_2\rangle =-\sin \xi |1\rangle + \cos \xi |2\rangle .
\een  
Thus, for the present case, $U_{11}=U_{22}=\cos \xi$ and $U_{21}=-U_{12}=\sin \xi$. }

Let us also assume that the spectral densities for sites 1 and 2 are the same, which we denote as ${\mathcal J}(\omega)$.  Namely,  ${\mathcal J}_{11}(\omega)={\mathcal J}_{22} (\omega)={\mathcal J} (\omega)$.  In addition, let us also assume that ${\mathcal J}_{12}(\omega)={\mathcal J}_{21}(\omega)=0$.  Then, ${\mathcal J}^{(1)}_{1,12}(\omega)={\mathcal J}^{(1)}_{2,21}(\omega)={\mathcal J} (\omega)$ and ${\mathcal J}^{(2)}_{12,12}(\omega)={\mathcal J}^{(2)}_{21,21}(\omega)=2{\mathcal J}(\omega)$.  For all other indices, ${\mathcal J}^{(1)}_{j,j'k'}(\omega)=0$ and ${\mathcal J}^{(2)}_{jk,j'k'}(\omega)=0$.

Then, all the bath correlation functions that enter the p-PQME can be represented by the following three functions:
\ben
&&{\mathcal K}(t)\equiv{\mathcal K}_{12,12}(t)=\frac{2}{\pi\hbar} \int_0^\infty d\omega \frac{{\mathcal J}(\omega)}{\omega^2} W_h(\omega)^2 \nonumber \\
&&\hspace{.5in}\times \left (\coth\left (\frac{\beta\hbar \omega}{2}\right) \cos (\omega t) - i \sin (\omega t) \right ) \nonumber \\
&&\hspace{.3in}={\mathcal K}_{21,21}(t)=-{\mathcal K}_{12,21}(t)=-{\mathcal K}_{21,12}(t) ,  \label{eq:kt-2}\\
&&{\mathcal M}(t)\equiv{\mathcal M}_{1,12}(t)=\frac{1}{\pi} \int_0^\infty d\omega \frac{{\mathcal J}(\omega)}{\omega} W_h(\omega) (1 -W_h(\omega)) \nonumber \\
&&\hspace{.5in} \times \left (\cos (\omega t)-i\coth \left (\frac{\beta\hbar\omega}{2}\right) \sin (\omega t) \right ) \nonumber \\ 
&&\hspace{.3in}={\mathcal M}_{2,21} (t)=-{\mathcal M}_{1,21}(t)=-{\mathcal M}_{2,12}(t) ,  \label{eq:mt-2}\\
&&{\mathcal C}(t)\equiv{\mathcal C}_{11}(t)=\frac{\hbar}{\pi} \int_0^\infty d\omega {\mathcal J} (\omega) (1-W_h(\omega))^2 \nonumber \\
&&\hspace{.5in} \times \left  ( \coth \left (\frac{\beta\hbar\omega}{2}\right ) \cos (\omega t)-i \sin (\omega t) \right ) \nonumber \\
&&\hspace{.3in}={\mathcal C}_{22}(t) . \label{eq:ct-2}
\een 
For all other components, ${\mathcal K}_{jk,j'k'}(\omega)=0$, ${\mathcal M}_{j,j'k'}(\omega)=0$, and ${\mathcal C}_{jk}(\omega)=0$. 

In the site basis,  Eq. (\ref{eq:qme_hom}) for the present case can be expressed as follows:
\ben
&&{\mathcal R}(t)\tilde \sigma_I(t) = \int_0^t d\tau \left \{ C(t-\tau) \left ([{\mathcal T}_{11}(t), {\mathcal T}_{11}(\tau) \tilde \sigma_I (t)] \right . \right .\nonumber \\
&&\left . \hspace{1.3 in}+[{\mathcal T}_{22}(t), {\mathcal T}_{22}(\tau) \tilde \sigma_I (t)] \right ) \nonumber \\
&&\hspace{.1in}+Jw {\mathcal M}(t-\tau) \left ( [\Delta {\mathcal T}_{d}(t), \Delta {\mathcal T}_{c}(\tau) \tilde \sigma_I(t)] \right .\nonumber \\
&&\hspace{0.9in}\left . -[\Delta {\mathcal T}_{c}(t), \Delta {\mathcal T}_{d}(\tau) \tilde \sigma_I(t)] \right ) \nonumber \\
&&\hspace{.1in}+J^2w^2 \left (e^{-{\mathcal K}(t-\tau)}-1 \right ) \left ([{\mathcal T}_{12}(t),{\mathcal T}_{12}(\tau) \tilde \sigma_I(t)] \right . \nonumber \\
&&\hspace{1.4in}\left . +[{\mathcal T}_{21}(t),{\mathcal T}_{21}(\tau) \tilde \sigma_I(t)] \right ) \nonumber \\  
&&\hspace{.1in}+J^2w^2 \left (e^{{\mathcal K}(t-\tau)}-1 \right ) \left ([{\mathcal T}_{12}(t),{\mathcal T}_{21}(\tau) \tilde \sigma_I(t)] \right . \nonumber \\
&&\hspace{1.4in}\left . \left . +[{\mathcal T}_{21}(t),{\mathcal T}_{12}(\tau) \tilde \sigma_I(t)] \right ) \right \} \nonumber \\
&&+{\rm H.c.} ,  \label{eq:rt-sigma-2state-1}
\een
where
\ben
\Delta {\mathcal T}_d(t)={\mathcal T}_{22}(t)-{\mathcal T}_{11}(t) ,\\
\Delta {\mathcal T}_c(t)={\mathcal T}_{21}(t)-{\mathcal T}_{12}(t) .
\een
Equation (\ref{eq:rt-sigma-2state-1}) above provides clear insights into the effects of p-PT and is useful for understanding the steady state behavior of the population dynamics.  However, for general numerical calculation, it is convenient to use expressions defined in the basis of eigenstates of $\tilde H_{0,s}$, as detailed in Sec. IIC.  For this, ${\mathcal R}^{p'q'}_{pq}(t)$ given by Eq. (\ref{eq:r-pq}) needs to be calculated, which in turn requires calculation of ${\mathcal W}_{jk,j'k'}^{pq}(t)$, ${\mathcal Y}_{j,j'k'}^{pq}(t)$, and ${\mathcal X}_{jj'}^{pq}(t)$ defined by Eqs. (\ref{eq:W-def})-(\ref{eq:X-def}).   For the present model of two state systems coupled to independent baths, most of these are zero except for few functions, as defined below.  

First, all of nonzero ${\mathcal W}_{jk,j'k'}^{pq}(t)$'s  defined by Eq. (\ref{eq:W-def}) for the present case reduce to one of the following two functions:
\ben
&&{\mathcal W}_{-}^{pq}(t)\equiv{\mathcal W}_{12,12}^{pq}(t)={\mathcal W}_{21,21}^{pq}(t)\nonumber \\
&&\hspace{.2in}= J^2w^2 \int_0^t d\tau e^{i\delta {\mathcal E}_{pq}(t-\tau)/\hbar} \left (e^{-{\mathcal K}(t-\tau)}-1 \right)  , \label{eq:w-pq}\\
&&{\mathcal W}_{+}^{pq}(t)\equiv{\mathcal W}_{12,21}^{pq}(t)={\mathcal W}_{21,12}^{pq}(t)\nonumber \\
&&\hspace{.2in}= J^2w^2 \int_0^t d\tau e^{i\delta {\mathcal E}_{pq}(t-\tau)/\hbar}\left (e^{{\mathcal K}(t-\tau)}-1 \right)  . \label{eq:w+pq}
\een
All other terms of ${\mathcal W}_{jk,j'k'}^{pq}(t)$ are zero by definition.  
Similarly, all of nonzero ${\mathcal Y}_{j,j'k'}^{pq}(t)$'s and ${\mathcal X}_{jj'}^{pq}(t)$'s  can be specified by
\ben
&&{\mathcal Y}^{pq}(t)\equiv{\mathcal Y}_{1,12}^{pq}(t)={\mathcal Y}_{2,21}^{pq}(t)\nonumber \\
&&\hspace{.5in}=Jw\int_0^t d\tau e^{i\delta {\mathcal E}_{pq}(t-\tau)/\hbar} {\mathcal M}(t-\tau)  \nonumber  \\
&&\hspace{.5in}=-{\mathcal Y}_{1,21}^{pq}(t)=-{\mathcal Y}_{2,12}^{pq}(t) ,  \label{eq:y-pq}\\
&&{\mathcal X}^{pq}(t)\equiv{\mathcal X}^{pq}_{11}(t)={\mathcal X}^{pq}_{22}(t)\nonumber \\
&&\hspace{.5in}=\int_0^t d\tau e^{i\delta {\mathcal E}_{pq}(t-\tau)/\hbar} {\mathcal C} (t-\tau) .
\een
All other terms with indices different from above are zero.   Thus, with ${\mathcal W}_{\pm}^{pq}(t)$, ${\mathcal Y}^{pq}(t)$, and ${\mathcal X}^{pq}(t)$ as defined above, all of ${\mathcal R}_{pq}^{p'q'}(t)$ constituting Eq. (\ref{eq:r-pq}) for the present case can be calculated.  

The first order inhomogeneous term, Eq. (\ref{eq:i1-exciton}), involves additional functions $f_{jk,k'}(t)$ and $h_{j,k'}(t)$, which can also be simplified for the present model. 
Let us define 
\ben
f(t)&\equiv& \exp\left \{\frac{2i}{\pi\hbar} \int_0^\infty d\omega \frac{{\mathcal J}(\omega)}{\omega^2} W_h(\omega)^2 \sin (\omega t)\right\}  \nonumber \\
&=&f_{12,1}(t) .
\een
Then, it is easy to show that $f_{21,2}(t)=f(t)$, $f_{21,1}(t)=f_{12,2}(t)=f^*(t)$.  On the other hand, $f_{11,1}(t)=f_{22,1}(t)=f_{11,2}(t)=f_{22,2}(t)=1$. 
For $h_{j,k'}(t)$, only specification of the following function is necessary.
\ben
h(t)&\equiv &\frac{2}{\pi}\int_0^\infty d\omega \frac{{\mathcal J}(\omega)}{\omega} (1-W_h(\omega)) W_h(\omega) \cos(\omega t) \nonumber \\ 
&=&h_{1,1}(t)=h_{2,2}(t)  .
\een
For other cases, $h_{1,2}(t)=h_{2,1}(t)=0$. 

\re{The second order inhomogeneous term, Eq. (\ref{eq:i2-exciton}), consists of much more terms that involve time integration of four different types of time correlation functions as described in Appendix C.   Although calculation of all these expressions is straightforward, actual numerical implementation is nontrivial and will be the subject of future work. } 

\subsection{Model calculations without inhomogeneous terms for Ohmic spectral density}
\re{Numerical calculations neglecting the contribution of inhomogeneous terms are provided here.  The objective of these calculations is to demonstrate} the feasibility of using the p-PQME expressions derived \re{in previous subsections} and \re{to investigate }the effects of the weighting function $W_h(\omega$).  \re{Thus, the results to be presented do not yet have the quantitative accuracy for all time due to the neglect of  inhomogeneous terms.  However, they still provide important qualitative information and reliable steady state limits for the cases where the inhomogeneous terms vanish.} 

\re{First, it is useful to further simplify the expressions for ${\mathcal R}_{pq}^{p'q'}(t)$'s.} Summing up all possible 16 cases of $j,k,j',k'=1,2$ in Eq. (\ref{eq:r-pq}), it is straightforward to show that it can be simplified as follows: 
\ben
{\mathcal R}_{pq}^{p'q'}(t)&=&\frac{1}{\hbar^2} \Bigg \{ \delta_{q'q} e^{i\delta {\mathcal E}_{pp'} t/\hbar} \sum_{r=1}^2 \left ( {\mathcal A}_{prrp'} {\mathcal X}^{p'r}(t) \right . \nonumber \\
&&\hspace{.3in}+( {\mathcal B}^{(1)}_{pr}{\mathcal B}^{(2)}_{rp'}-{\mathcal B}_{pr}^{(2)}{\mathcal B}_{rp'}^{(1)} ) {\mathcal Y}^{p'r} (t) \nonumber \\
&&\hspace{.3in}\left . +{\mathcal C}_{prrp'}^{(1)} {\mathcal W}_{-}^{p'r}(t) +{\mathcal C}_{prrp'}^{(2)} {\mathcal W}_{+}^{p'r}(t) \right ) \nonumber \\
&& \hspace{.1in}- e^{i(\delta {\mathcal E}_{pp'}-\delta {\mathcal E}_{qq'})t/\hbar} \left ( {\mathcal A}_{q'qpp'} {\mathcal X}^{p'p}(t) \right .\nonumber \\
&& \hspace{.3in} +({\mathcal B}_{q'q}^{(1)} {\mathcal B}_{pp'}^{(2)}-{\mathcal B}_{q'q}^{(2)} {\mathcal B}_{pp'}^{(1)} ) {\mathcal Y}^{p'p} (t) \nonumber \\
&&\hspace{.3in} \left . +{\mathcal C}_{q'qpp'}^{(1)} {\mathcal W}_{-}^{p'p}(t)+{\mathcal C}_{q'qpp'}^{(2)} {\mathcal W}_{+}^{p'p}(t)\right ) \Bigg\} \nonumber \\
&&\hspace{.5in}+[{\rm C.C.}, p\leftrightarrow q, p'\leftrightarrow q'], 
\een
where 
\ben
&&{\mathcal A}_{pqrs}=U_{1p}U_{1q}U_{1r}U_{1s}+U_{2p}U_{2q}U_{2r}U_{2s} , \\
&&{\mathcal B}^{(1)}_{pq}=U_{1p}U_{1q}-U_{2p}U_{2q} , \\
&&{\mathcal B}^{(2)}_{pq}=U_{1p}U_{2q}-U_{2p}U_{1q} , \\
&&{\mathcal C}^{(1)}_{pqrs}=U_{1p}U_{2q}U_{1r}U_{2s}+U_{2p}U_{1q}U_{2r}U_{1s} , \\
&&{\mathcal C}^{(2)}_{pqrs}=U_{1p}U_{2q}U_{2r}U_{1s}+U_{2p}U_{1q}U_{1r}U_{2s}  .
\een

The bath spectral density being considered is the Ohmic form with exponential cutoff as follows:
\be
{\mathcal J}(\omega)=\pi \hbar\eta\omega e^{-\omega/\omega_c} .
\ee
\re{It is assumed that the two parameters of the above spectral density are  set to $\eta=1$ and $\hbar\omega_c=200\ {\rm cm^{-1}}$.  In addition, the temperature and electronic coupling are set to $T=300\ {\rm K}$ and $J=300\ {\rm cm^{-1}}$. For this choice of parameters, the bath dynamics, system electronic coupling, and thermal energy are all comparable.  Thus, the dynamics are expected to be the border-line case between incoherent and coherent quantum dynamics. Two cases of relative site energies, $E_1=E_2$ and $E_1-E_2=200\ {\rm cm^{-1}}$ will be considered.   }

\begin{figure}
\includegraphics[width=3.in]{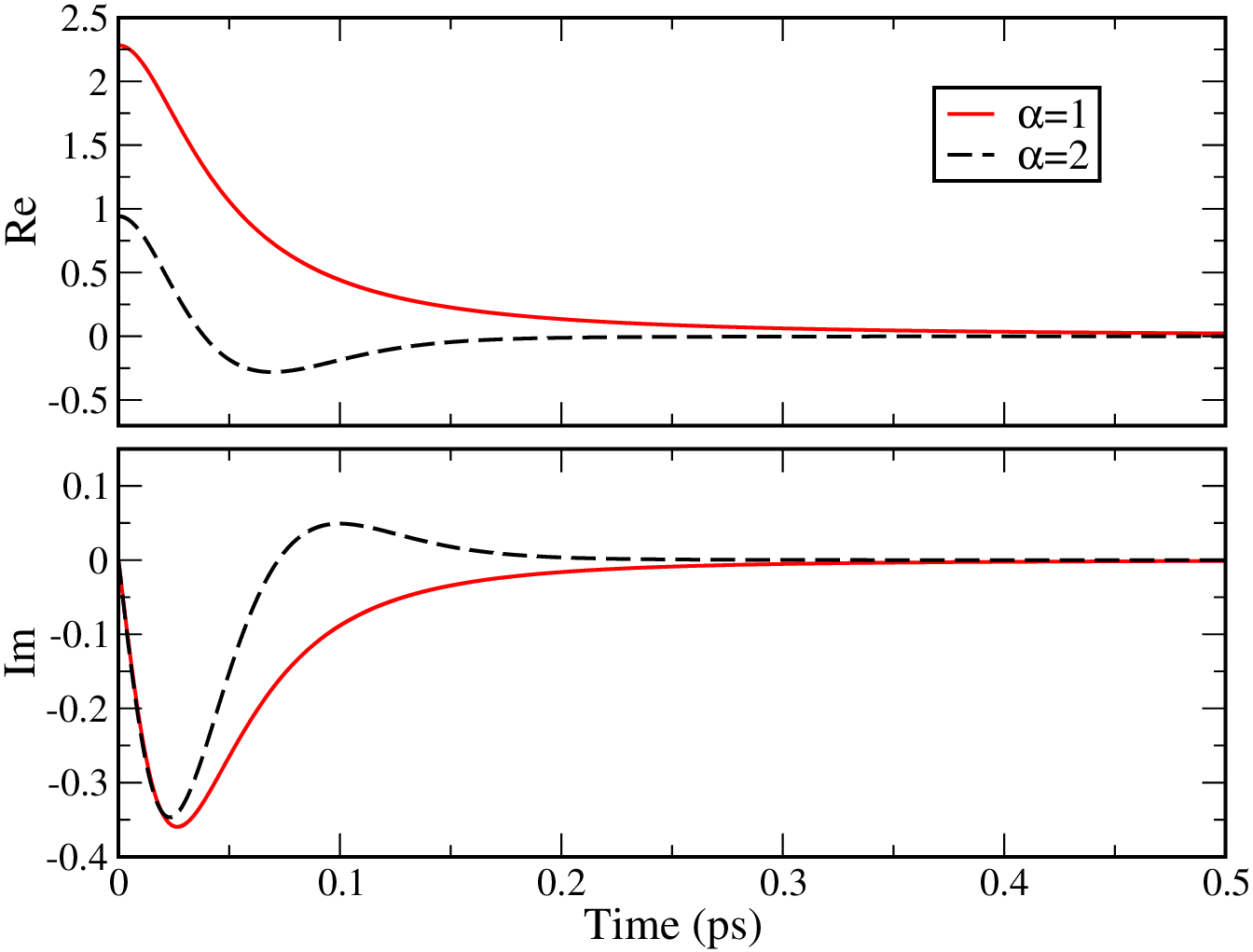}
\caption{\re{Real (upper panel) and imaginary (Lower panel) parts of ${\mathcal K}(t)$ for $\omega_h/\omega_c=1$ and for two cases of $\alpha=1$ and $2$.  \label{fig1}}}
\end{figure}

There are infinite number of possible choices available for $W_h(\omega)$ that satisfies the requirement of Eq. (\ref{eq:property-wh}). 
\re{One simple but flexible choice is the following function known as (cumulative) Weibull distribution,\cite{weibull-jam18} or the complement of a stretched exponential function: }
\be
W_h(\omega)= 1 - e^{-(\omega/\omega_h)^\alpha} . \label{eq:wh-alpha}
\ee
\re{For small $\omega$, $W_h(\omega) \approx (\omega/\omega_h)^\alpha$.  For large $\alpha$, $W_h(\omega)$ approaches the step function at $\omega=\omega_h$. }

\re{While choosing a value $\alpha$ in Eq. (\ref{eq:wh-alpha}) such that $\alpha \geq 1$ is sufficient for ensuring that the Debye-Waller factor does not vanish, it is not clear whether it also guarantees well-behaving time evolution equation.   In order to check this, it is important to examine how the three time correlation functions defined by Eqs. (\ref{eq:kt})-(\ref{eq:ct}) behave for different choices of $\alpha$.   Figure \ref{fig1} shows the real and imaginary parts of ${\mathcal K}(t)$ defined by Eq. (\ref{eq:kt}) for two different values of $\alpha=1$ and $2$ with the choice of $\omega_h=\omega_c$.  In both cases, the real and imaginary parts decay to zero quickly enough to make both ${\mathcal W}_{-}^{pq}(t)$ and ${\mathcal W}_{+}^{pq}(t)$, defined respectively by Eqs. (\ref{eq:w-pq}) and (\ref{eq:w+pq}), converge to finite values.  }

\begin{figure}
\includegraphics[width=3.in]{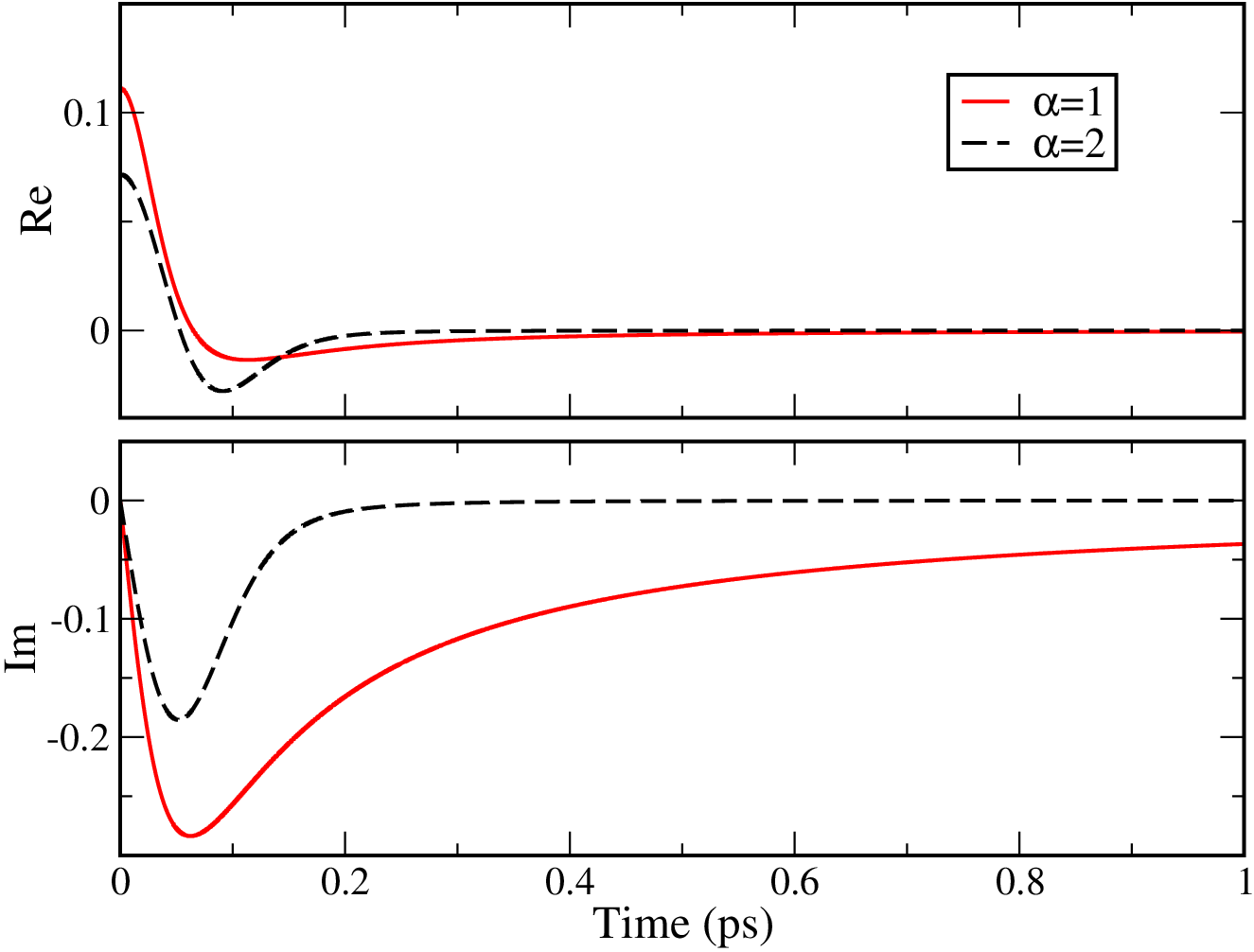}
\caption{\re{Real (upper panel) and imaginary (lower panel) parts of ${\mathcal M}(t)/(\hbar\omega_c)$ for  $\omega_h/\omega_c=1$ and for two cases of $\alpha=1$ and $2$. \label{fig2}}}
\end{figure}

\re{On the other hand, for the case of ${\mathcal M(t)}$, it turns out that the choice of $\alpha=1$ does not result in a stable time evolution equation in general.  Figure  \ref{fig2} shows the real and imaginary parts of ${\mathcal M}(t)$, also for $\alpha=1$ and $2$ with the choice of $\omega_h=\omega_c$.   For $\alpha=1$, while the real part decays to zero quickly, the imaginary part is seen to decay very slowly.  In fact, due to the slow, the ${\mathcal Y}^{pq}(t)$ defined by Eq. (\ref{eq:y-pq}) diverges in general in this case.  Test calculations for other values of $\alpha$ show that such divergence persists up to $\alpha=3/2$.   On the other hand, for the case of $\alpha=2$ shown in Fig. \ref{fig2}, it is clear that the imaginary part decays to zero quickly, ensuring for ${\mathcal Y}^{pq}(t)$ to converge to a finite value in the steady state limit. }

\begin{figure}
\includegraphics[width=3.in]{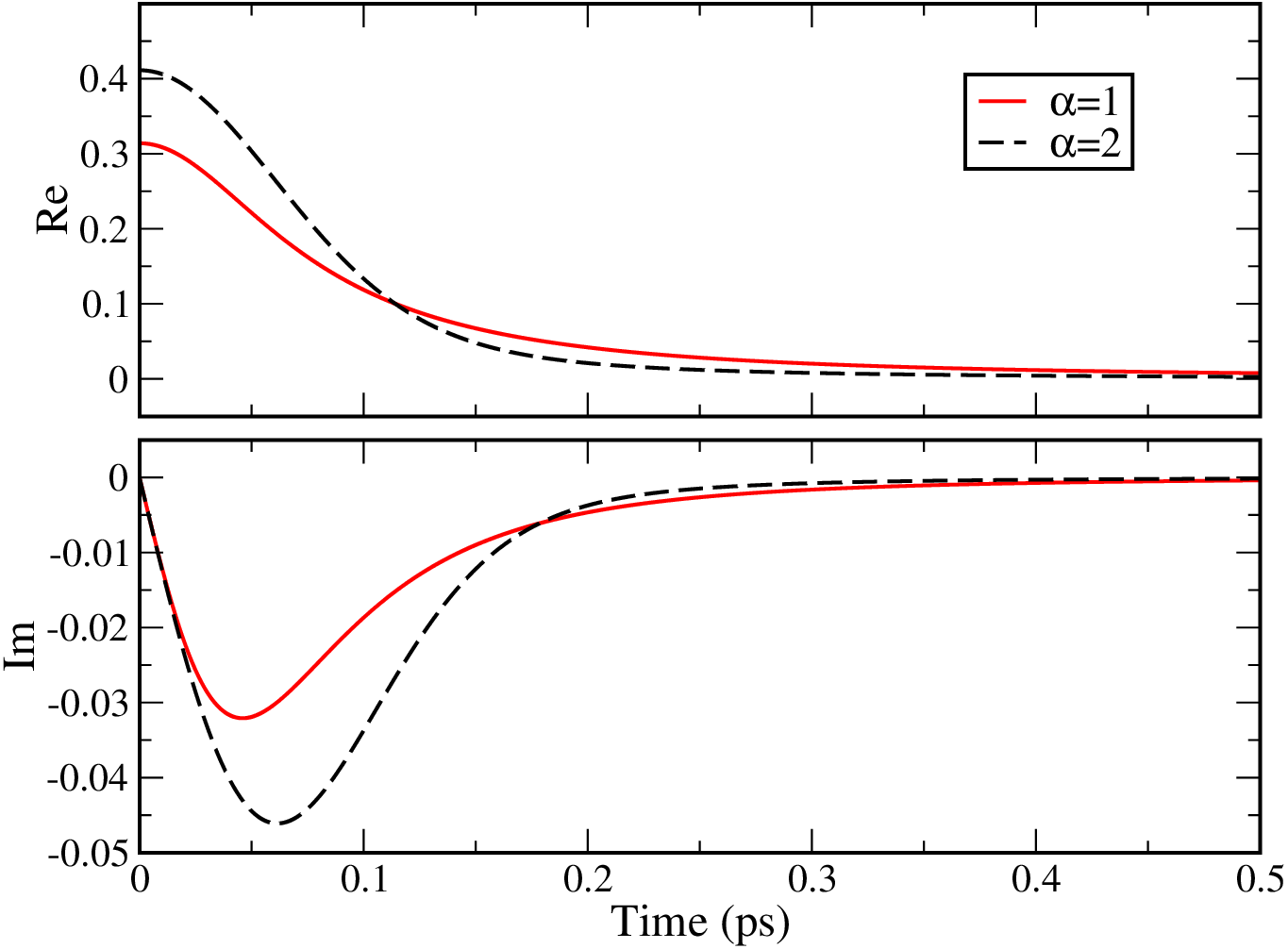}
\caption{\re{Real (upper panel) and imaginary (lower panel) parts of ${\mathcal C}(t)/(\hbar^2\omega_c^2)$ fo
r  $\omega_h/\omega_c=1$ and for two cases of $\alpha=1$ and $2$. \label{fig3}}}
\end{figure}

\re{For the case of ${\mathcal C}(t)$, as shown in Fig. \ref{fig3}, both real and imaginary parts decay to zero quickly already for $\alpha=1$, resulting in well behaving ${\mathcal X}^{pq}(t)$.   Summing up the results shown in Figs. \ref{fig1}-\ref{fig3}, while the choice of $\alpha=1$ is acceptable as far as the Deby-Waller factor, ${\mathcal K}(t)$, and ${\mathcal C}(t)$ are concerned, it is not appropriate due to resulting slow decay of ${\mathcal M}(t)$.   On the other hand, the choice of $\alpha=2$ results in all well-behaving and convergent functions that constitute the relaxation operator.  This is also true for all the terms involved in the inhomogeneous terms as well.   }  

\begin{figure}
\includegraphics[width=3.2in]{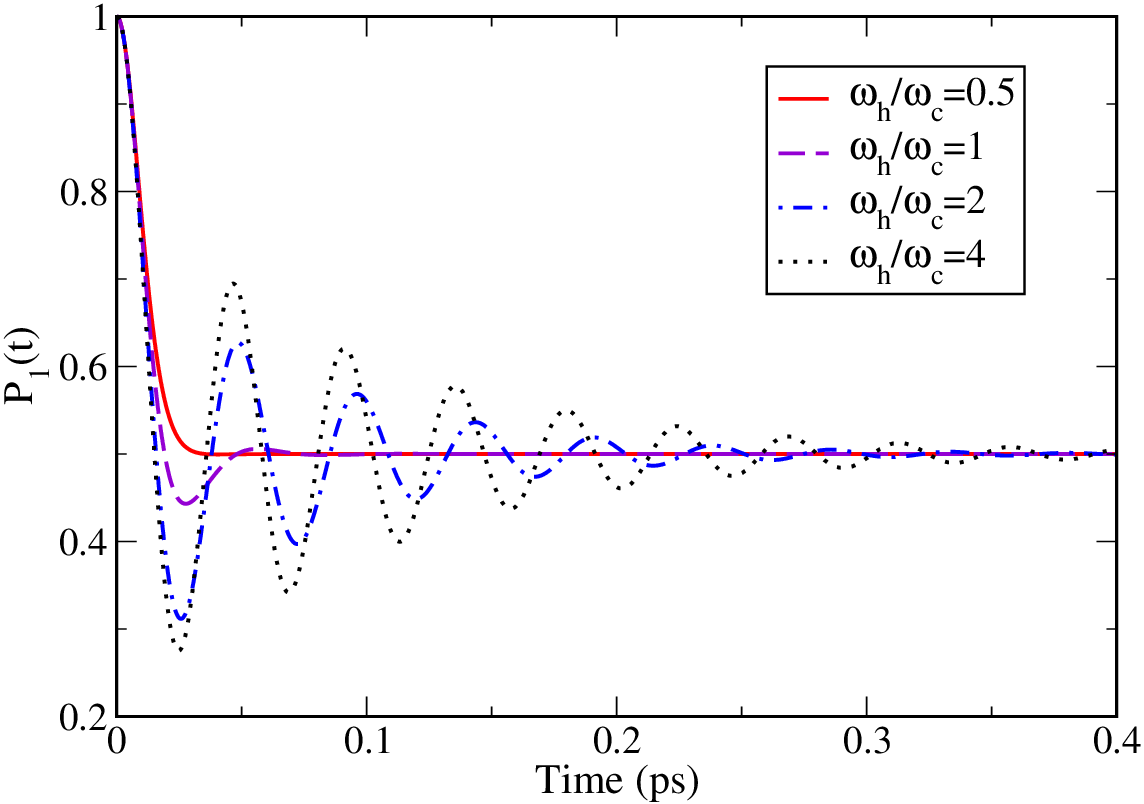}
\caption{\re{Populations of donor (site $1$) for different values of $\omega_h$ for $E_1=E_2$ and $\alpha=2$ in $W_h(\omega)$, Eq. (\ref{eq:wh-alpha}).  Other parameters of the model are as follows: $\eta=1$, $\omega_c=200\ {\rm cm^{-1}}$, $T=300\ {\rm K}$, and $J=300\ {\rm cm^{-1}}$. \label{fig4} }}
\end{figure}

\re{Figure \ref{fig4} shows results for $E_1=E_2$ for different values of $\omega_h$ with the choice of $\alpha=2$, where $W_h(\omega)$ becomes a complement of a Gaussian function. The result for the smallest value of $\omega_h$ among those shown ($\omega_h/\omega_c=0.5$) is close to the limit of full PQME, without coherence, whereas  that for  the largest value of $\omega_h$ among those shown ($\omega_h/\omega_c=4$) is close to the full 2nd order time-local QME, which has maximum coherence.  For $\omega_h/\omega_c=1$, the time dependent population exhibits an intermediate character between the two limits.} 

\begin{figure}
\includegraphics[width=3.2in]{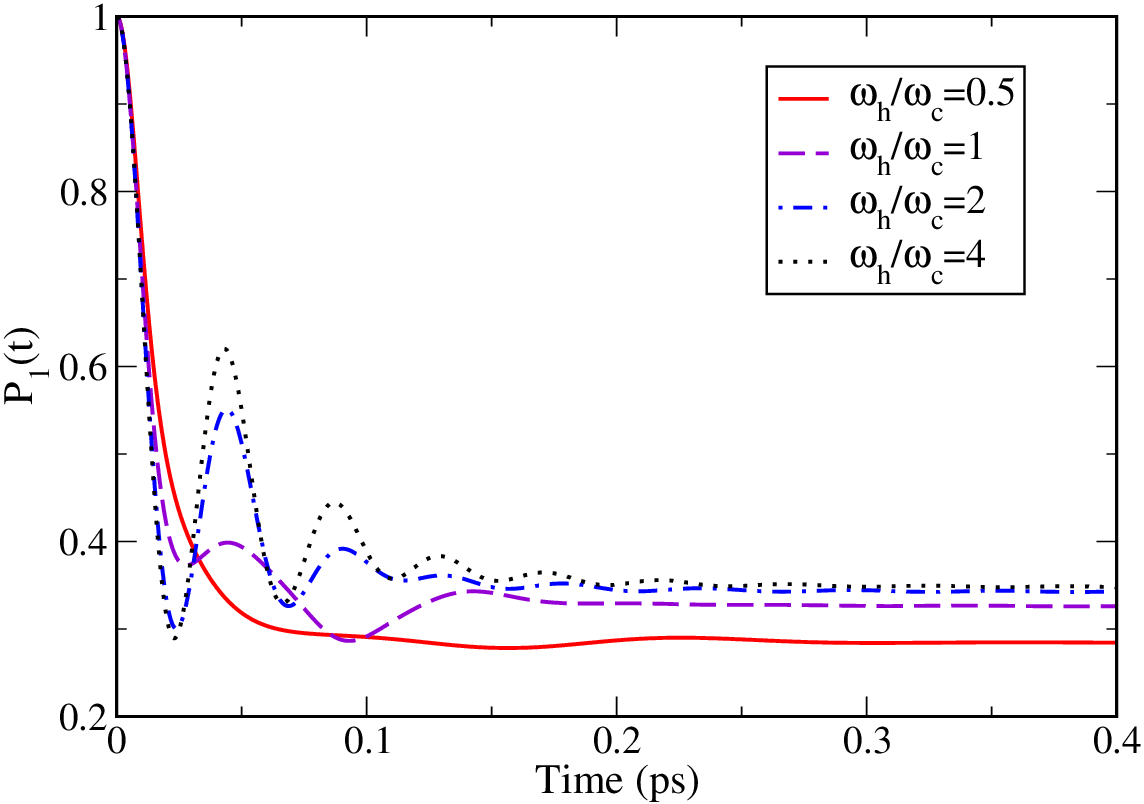}
\caption{\re{Populations of donor (site $1$) for $E_1-E_2=200\ {\rm cm^{-1}}$ and $\alpha=2$ in $W_h(\omega)$, Eq. (\ref{eq:wh-alpha}).   All other parameters are the same as in Fig. 4.\label{fig5}} }
\end{figure}

\re{Figure \ref{fig5} shows results for an asymmetric case, where $E_1-E_2=200 \ {\rm cm^{-1}}$.   All other parameters, including $\alpha=2$, remain the same as those for Fig. \ref{fig4}.  In this case, the steady state limits of population as well as the coherence pattern vary with $\omega_h$.   The variation of the steady state limit with $\omega_h$ reflects different extent of system-bath entanglement depending on the extent of polaron transformation. The smaller the value of $\omega_h$, the closer the steady states are to the original localized states $1$ and $2$, for which the energy gap becomes the maximum. }

\section{Concluding Remarks} 
This work has provided a general framework to overcome a known issue of the original 2nd order PQME, namely, premature over-relaxation of the sluggish bath, by deriving full expressions for the second order time-local p-PQME.  The main results of this work, represented by Eqs. (\ref{eq:r-pq}), (\ref{eq:i1-exciton}), and (\ref{eq:i2-exciton}), can be applied for any kinds of bath spectral densities and initial system states but will be particularly useful for the cases where the bath spectral densities are Ohmic or sub-Ohmic.  \re{It is important to note that the expressions provided here are applicable even to the case where the same bath mode is partly transformed, with the remaining part untransformed, and is, thus, more general than the case where
the bath is divided into two disjoint transformed and untransformed groups. }

Numerical tests for a simple two level system coupled to an Ohmic bath demonstrate that appropriate \re{specification} of the weighting function $W_h(\omega)$ \re{can tune the extent of coherence and the extent of system-bath entanglement in the steady state limit.  This adds} a new dimension of flexibility in incorporating the PT approach into a QME calculation.
The flexibility in choosing the weighting function $W_h(\omega)$ in \re{all the expressions derived }for the p-PQME presented here leaves open various possibilities of adapting or improving the methodology.  For example, variational theorem can be used for its optimization.  Alternatively, benchmarking against numerically exact computational results followed by a Machine Learning based optimization can potentially lead to an optimized second order p-PQME that can best approximate exact dynamics.  To this end, full calculations including all the inhomogeneous terms and benchmarking against a broad range of exact numerical results, the subject of a forthcoming work, will be necessary.

The formulation developed here also will be useful for further extension of PT based QME approaches.  For example, extension to the cases with time dependent Hamiltonian for driven quantum systems is straightforward.  The formulations and theoretical identities employed in this work will also be useful for the development of new PT based approaches for general anharmonic bath and for the formulation of time dependent PT approach.          

\acknowledgments

This work was mainly supported by the National Science Foundation (CHE-1900170).  The author also acknowledges partial support from the US Department of Energy, Office of Sciences, Office of Basic Energy Sciences (DE-SC0021413) and support from Korea Institute for Advanced Study (KIAS) through its KIAS Scholar program. \vspace{.2in}\\

\noindent
{\bf  AUTHOR DECLARATIONS} \vspace{.1in}\\
{\bf Conflict of Interest} \vspace{.1in}\\
The author has no conflicts to disclose. \vspace{.2in}\\
\noindent
{\bf  DATA AVAILABILITY} \vspace{.1in}\\
Most data that support the findings of this article are contained in this article.  Additional data are available from the corresponding author upon reasonable request.

\appendix
\section{Derivation of Eq. (\ref{eq:blm_eq})}
The trace of $\tilde B_{jk}(t)\tilde B_{j'k'}(\tau)$ with $\rho_b$ in Eq. (\ref{eq:qme_hom}), namely, its thermal average is expressed as follows:
\ben
&&\langle \tilde B_{jk}(t)\tilde B_{j'k'}(\tau)\rangle=\langle \theta_j^\dagger (t)\theta_k(t)\theta_{j'}^\dagger (\tau)\theta_{k'}(\tau)\rangle \nonumber \\
&&\hspace{.3in}-w_{jk}\langle \theta_{j'}^\dagger(\tau)\theta_{k'}(\tau)\rangle-w_{j'k'}\langle \theta_j^\dagger(t)\theta_{k}(t)\rangle+w_{jk}w_{j'k'}\nonumber \\
&&\hspace{.3in}+\delta_{jk}\left (\langle D_j(t)\theta_{j'}^\dagger(\tau)\theta_{k'}(\tau)\rangle - \langle D_j(t)\rangle w_{j'k'} \right) \nonumber \\
&&\hspace{.3in}+\delta_{j'k'}\left (\langle \theta_j^\dagger(t)\theta_{k}(t)D_{j'}(\tau)\rangle -w_{jk}\langle D_{j'} (\tau)\rangle \right)\nonumber \\
&&\hspace{.3in}+\delta_{jk}\delta_{j'k'}\langle D_j(t)D_{j'}(\tau)\rangle \nonumber \\
&&=\langle \theta_j^\dagger (t)\theta_k(t)\theta_{j'}^\dagger (\tau)\theta_{k'}(\tau)\rangle-w_{jk}w_{j'k'} \nonumber \\
&&\hspace{.3in}+\delta_{jk} \langle D_j(t)\theta_{j'}^\dagger (\tau)\theta_{k'}(\tau)\rangle \nonumber \\
&&\hspace{.3in}+\delta_{j'k'}\langle \theta_j^\dagger(t)\theta_{k}(t) D_{j'}(\tau)\rangle \nonumber \\
&&\hspace{.3in}+\delta_{jk}\delta_{j'k'}\langle D_j(t)D_{j'}(\tau)\rangle . \label{eq:bb-cor}
\een
In obtaining the second equality of the above equation, the identities that $\langle \theta_{j'}^\dagger(\tau)\theta_{k'}(\tau)\rangle=w_{j'k'}$, $\langle \theta_j^\dagger(t)\theta_{k}(t)\rangle=w_{jk}$, and $\langle D_{j} (t)\rangle=\langle D_{j'} (\tau)\rangle=0$ have been used.

The first term in the second equality of Eq. (\ref{eq:bb-cor}) has the same form as that for the full PT\cite{jang-jcp131} except for the additional factor $W_h(\omega_n)$ multiplied with $\delta g_{n,jk}$ and $\delta g_{n,j'k'}$, respectively.    Thus, following the same procedure as in previous work,\cite{jang-jcp131} it can be shown to be
\be
\langle \theta_j^\dagger (t)\theta_k(t)\theta_{j'}^\dagger (\tau)\theta_{k'}(\tau)\rangle =w_{jk}w_{j'k'} e^{-{\mathcal K}_{jk,j'k'} (t-\tau)} , \label{eq:theta-4}
\ee
where ${\mathcal K}_{jk,j'k'} (t-\tau)$ in the exponent  is defined by Eq. (\ref{eq:kt}).
  
The third term in the second equality of Eq. (\ref{eq:bb-cor}) can be expressed as 
\ben
&&\langle D_j(t)\theta_{j'}^\dagger (\tau)\theta_{k'}(\tau)\rangle \nonumber \\
&&=\sum_n \hbar\omega_n g_{n,j}(1-W_h(\omega_n))\Big \langle (b_n e^{-i\omega_n t}+b_n^\dagger e^{i\omega_n t})  \nonumber \\
&&\hspace{.2in}\times \prod_{n'} e^{\delta g_{n',j'k'} W_h(\omega_{n'}) (b_{n'}^\dagger e^{i\omega_{n'} \tau} -b_{n'} e^{-i\omega_{n'} \tau  } )} \Big \rangle .  \label{eq:djt} 
\een
In the above expression, the product of bath modes (indexed by $n'$) can be averaged independently for $n'\neq n$, resulting in  
\ben
&&\left \langle \prod_{n'\neq n} e^{\delta g_{n',j'k'} W_h(\omega_{n'}) (b_{n'}^\dagger e^{i\omega_{n'} \tau} -b_{n'} e^{-i\omega_{n'} \tau  } )} \right \rangle \nonumber \\
&&=\prod_{n'\neq n} e^{-\coth(\beta\hbar\omega_{n'}/2)\delta g_{n',j'k'}^2 W_h(\omega_{n'})^2/2} .
\een
On the other hand,  the term for $n'=n$ in Eq. (\ref{eq:djt}) need to be calculated together with the linear term as follows: 
\ben
&&\langle (b_ne^{-i\omega_n t}+b_n^\dagger e^{i\omega_n t}) e^{\delta g_{n,j'k'} W_h(\omega_n) (b_n^\dagger e^{i\omega_n \tau} -b_n e^{-i\omega_n\tau})}\rangle  \nonumber \\
 &&= e^{-i\omega_n t} \langle b_n e^{\gamma^* b_n^\dagger-\gamma b_n}\rangle +e^{i\omega_n t} \langle b_n^\dagger e^{\gamma^* b_n^\dagger-\gamma b_n}\rangle , \label{eq:b-gamma}
 \een 
 where $\gamma$ is defined by
 \be
 \gamma=\delta g_{n,j'k'} W_h(\omega_{n}) e^{-i\omega_{n} \tau} , \label{eq:def-gam}
 \ee  
 and $\gamma^*$ is the complex conjugate of $\gamma$. 
 Using the fact that $e^{\gamma^* b_n^\dagger-\gamma b_n}=e^{-\gamma b_n} e^{\gamma^* b_n^\dagger}e^{|\gamma|^2/2}$ 
 and the following identity:
 \ben
 &&\langle b_n e^{-\gamma b_n} e^{\gamma^* b_n^\dagger}\rangle=-\frac{\partial}{\partial \gamma} \langle  e^{-\gamma b_n} e^{\gamma^* b_n^\dagger}\rangle \nonumber \\
&&\hspace{.5in}=\frac{\gamma^*}{1-e^{-\beta\hbar\omega_n} }e^{-\gamma\gamma^*/(1-e^{-\beta\hbar\omega_n})}  ,  \label{eq:b-ebb}
\een
it is straightforward to show that 
\be
\langle b_n e^{\gamma^* b_n^\dagger-\gamma b_n}\rangle=\frac{\gamma^*}{1-e^{-\beta\hbar\omega_n} }e^{-\coth(\beta\hbar\omega_n/2)|\gamma|^2/2} .\label{eq:b1}
\ee

\re{Similarly, the bath average in the second term of Eq. (\ref{eq:b-gamma}) can be expressed as:
\be
\langle b_n^\dagger e^{\gamma^* b_n^\dagger -\gamma b_n} \rangle = \langle b_n^\dagger e^{-\gamma b_n}e^{\gamma^* b_n^\dagger} \rangle e^{|\gamma|^2/2} . \label{eq:bd-gamma}
\ee 
Using the following identity:
\be 
b_n^\dagger e^{-\gamma b_n}=e^{-\gamma b_n} b_n^\dagger +\gamma e^{-\gamma b_n} ,
\ee
the average over the bath in Eq. (\ref{eq:bd-gamma}) can be expressed as 
\be
\langle b_n^\dagger e^{-\gamma b_n} e^{\gamma^* b_n^\dagger} \rangle = \langle e^{-\gamma b_n} b_n^\dagger e^{\gamma^* b_n^\dagger}\rangle +\gamma \langle e^{-\gamma b_n} e^{\gamma^* b_n^\dagger} \rangle  \label{eq:bd-gamma-2}
\ee
Now, employing the following identity
\ben
&&\langle e^{-\gamma b_n} b_n^\dagger e^{\gamma^* b_n^\dagger}\rangle=\frac{\partial}{\partial \gamma^*} \langle  e^{-\gamma b_n} e^{\gamma^* b_n^\dagger}\rangle \nonumber \\
&&\hspace{.5in}=-\frac{\gamma}{1-e^{-\beta\hbar\omega_n} }e^{-\gamma\gamma^*/(1-e^{-\beta\hbar\omega_n})}  , \label{eq:bd-ebb} 
\een
in Eq. (\ref{eq:bd-gamma-2}) and combining the resulting expression with Eqs. (\ref{eq:bd-gamma}), we obtain the following expression: }
\be
\langle b_n^\dagger e^{\gamma^* b_n^\dagger-\gamma b_n}\rangle=-\frac{\gamma e^{-\beta\hbar\omega_n}}{1-e^{-\beta\hbar\omega_n} }e^{-\coth(\beta\hbar\omega_n/2)|\gamma|^2/2} .\label{eq:b2}
\ee
Combining Eqs. (\ref{eq:b1}) and (\ref{eq:b2}) with the definition of Eq. (\ref{eq:def-gam}) leads to 
\ben
&&\langle (b_ne^{-i\omega_n t}+b_n^\dagger e^{i\omega_n t})e^{\delta g_{n,j'k'} W_h(\omega_n) (b_n^\dagger e^{i\omega_n \tau} -b_n e^{-i\omega_n \tau})} \rangle  \nonumber \\
 &&=\delta g_{n,j'k'} W_h(\omega_n)e^{-\coth(\beta\hbar\omega_{n}/2)\delta g_{n,j'k'}^2 W_h(\omega_{n})^2/2} \nonumber \\
 &&\times \left ( \cos (\omega_n (t-\tau)) -i\coth\left (\frac{\beta\hbar\omega_n}{2}\right) \sin (\omega_n (t-\tau)) \right) .\nonumber \\
\een
Employing this identity in Eq. (\ref{eq:djt}), one can find that 
\ben
&&\langle D_j(t)\theta_{j'}^\dagger (\tau)\theta_{k'}(\tau)\rangle \nonumber \\
&&=w_{j'k'}\sum_n \hbar\omega_n g_{n,j}(1-W_h(\omega_n)) \delta g_{n,j'k'}W_h(\omega_n) \nonumber \\ 
&&\hspace{.2in} \times  \left ( \cos (\omega_n (t-\tau)) -i\coth\left (\frac{\beta\hbar\omega_n}{2}\right) \sin (\omega_n (t-\tau)) \right) . \nonumber \\  \label{eq:djt-1} 
\een

The fourth term in the second equality of Eq. (\ref{eq:bb-cor}) can be expressed as 
\ben
&&\langle \theta_{j}^\dagger (t)\theta_{k}(t) D_{j'}(\tau)\rangle \nonumber \\
&&=\sum_n \hbar\omega_n g_{n,j'}(1-W_h(\omega_n)) \nonumber \\
&&\hspace{.2in}\times \Big \langle \prod_{n'} e^{\delta g_{n',jk} W_h(\omega_{n'}) (b_{n'}^\dagger e^{i\omega_{n'} t} -b_{n'} e^{-i\omega_{n'} t  } )} \nonumber \\
&&\hspace{.3in}\times (b_n e^{-i\omega_n \tau}+b_n^\dagger e^{i\omega_n \tau})  \Big \rangle .  \label{eq:b3}
\een
The \re{bath average term in} the above expression can be calculated in a manner similar to \re{the third term of Eq. (\ref{eq:bb-cor}), which has been described above,} but with a different definition of 
$\gamma=\delta g_{n,jk} W_h(\omega_n) e^{-i\omega_n t}$.  
\re{For this,  $\langle e^{\gamma^*b_n^\dagger-\gamma b_n} b_n \rangle $ and $\langle e^{\gamma^*b_n^\dagger-\gamma b_n} b_n^\dagger \rangle $ need to be calculated.  For the first of these terms,  the following identity can be used.}
\be
e^{\gamma^*b_n^\dagger} b_n =(b_n-\gamma^*)e^{\gamma^*b_n^\dagger}.
\ee
\re{Thus, 
\ben
&&\langle e^{\gamma^* b_n^\dagger -\gamma b_n} b_n\rangle=\langle e^{-\gamma b_n} e^{\gamma^* b_n^\dagger} b_n\rangle e^{|\gamma|^2/2}  \nonumber \\
&&=\langle e^{-\gamma b_n} \left (b_n e^{\gamma^* b_n^\dagger}-\gamma^* e^{\gamma b_n^\dagger}\right)\rangle e^{|\gamma|^2/2} \nonumber \\
&&=\left (\langle b_n e^{-\gamma b_n} e^{\gamma^* b_n^\dagger}\rangle -\gamma^* \langle e^{-\gamma b_n}e^{\gamma b_n^\dagger}\rangle \right) e^{|\gamma|^2/2} .
\een
Employing Eq. (\ref{eq:b-ebb}), one can simplify the above expression as follows:}
\be
\langle e^{\gamma^* b_n^\dagger-\gamma b_n} b_n\rangle=\gamma^* \frac{e^{-\beta\hbar\omega_n}}{1-e^{-\beta\hbar\omega_n}} e^{-|\gamma|^2 \coth(\beta\hbar\omega_n/2) } .
\ee
Combining the above identity with Eq. (\ref{eq:bd-ebb}) \re{followed by further calculation} leads to the following expression:
\ben
&&\langle  e^{\delta g_{n,jk} W_h(\omega_n) (b_n^\dagger e^{i\omega_n t} -b_n e^{-i\omega_n t})}(b_ne^{-i\omega_n \tau}+b_n^\dagger e^{i\omega_n \tau})\rangle  \nonumber \\
 &&=\delta g_{n,kj} W_h(\omega_n)e^{-\coth(\beta\hbar\omega_{n}/2)\delta g_{n,jk}^2 W_h(\omega_{n})^2/2} \nonumber \\
 &&\times \left ( \cos (\omega_n (t-\tau)) -i\coth\left (\frac{\beta\hbar\omega_n}{2}\right) \sin (\omega_n (t-\tau)) \right) ,\nonumber \\
\een
where note the use of the fact that $\delta g_{n,kj}=-\delta g_{n,jk}$ on the righthand side of the above equation. 
As a result,  Eq. (\ref{eq:b3}) can be expressed as
\ben
&&\langle \theta_{j}^\dagger (t)\theta_{k}(t) D_{j'}(\tau)\rangle \nonumber \\
&&=w_{jk}\sum_n \hbar\omega_n g_{n,j'}(1-W_h(\omega_n))\delta g_{n,kj} W_h(\omega_n) \nonumber \\
&&\times \left ( \cos (\omega_n (t-\tau)) -i\coth\left (\frac{\beta\hbar\omega_n}{2}\right) \sin (\omega_n (t-\tau)) \right) . \nonumber \\  \label{eq:b3-1}
\een

Calculation of the last term in the second equality of Eq. (\ref{eq:bb-cor}) is straightforward.  The resulting expression is as follows:
\ben
&&\langle D_j(t)D_{j'}(\tau)\rangle=\sum_n\hbar^2\omega_n^2 g_{n,j}g_{n,j'} (1-W_h(\omega_n))^2 \nonumber \\
&&\hspace{.5in}\times ( \langle b_n b_n^\dagger \rangle e^{-i\omega_n (t-\tau)}+ \langle b_n^\dagger b_n\rangle e^{i\omega_n (t-\tau)} ) \nonumber \\
&=&\sum_n\hbar^2\omega_n^2 g_{n,j}g_{n,j'} (1-W_h(\omega_n))^2 \nonumber \\
&&\times \left ( \coth \left (\frac{\beta\hbar\omega_n}{2}\right ) \cos (\omega_n (t-\tau)) -i \sin (\omega_n (t-\tau))\right ) . \nonumber\\ \label{eq:dd-t}
\een
In the second equality of the above equation, the fact that $\langle b_n^\dagger b_n\rangle=\langle b_n b_n^\dagger\rangle -1=e^{-\beta\hbar \omega_n}/(1-e^{-\beta\hbar\omega_n})$ has been used.

\section{Evaluation of the bath correlation function in the first order inhomogeneous term}
The bath portion of the first order inhomogeneous term, Eq. (\ref{eq:ihom-1}), can be expressed as follows:
\ben
&&Tr_b\{ \tilde B_{jk} (t)\delta \tilde \rho_{b,j'k'} \}=J_{jk} Tr_b \left \{ \theta_{k'} \theta_j^\dagger (t) \theta_k(t) \theta_{j'}^\dagger \rho_b \right\}\nonumber \\
&& \hspace{.5in}-J_{jk} w_{jk} w_{j'k'} +\delta_{jk} Tr_b\left\{\theta_{k'} D_j(t) \theta_{j'}^\dagger \rho_{b} \right\} . \label{eq:b-deltarho}
\een
In the first term on the righthand side of the above expression, the product of the first three operators within the trace operation can be expressed as 
\ben
&&\theta_{k'}\theta_j^\dagger(t)\theta_{k}(t)= \theta_{k'}\theta_j^\dagger(t)\theta_{k}(t)\theta_{k'}^\dagger\theta_{k'} \nonumber \\
&&=e^{\sum _n \delta g_{n,jk} W_h(\omega_n) (\theta_{k'} b_n^\dagger \theta_{k'}^\dagger e^{i\omega_nt} - \theta_{k'} b_n \theta_{k'}^\dagger e^{-i\omega_nt} )} \theta_{k'}  .\nonumber \\ \label{eq:bcor1-app-b}
\een 
In the exponent on the righthand side of the above expression, 
\ben
\theta_{k'} b_n^\dagger \theta_{k'}^\dagger&=&b_n^\dagger-W_h(\omega_n)g_{n,k'}[b_n^\dagger-b_n,b_n^\dagger] \nonumber \\
&=&b_n^\dagger+W_h(\omega_n)g_{n,k'} ,  \label{eq:theta-bnd-com}\\
\theta_{k'} b_n \theta_{k'}^\dagger&=&b_n-W_h(\omega_n)g_{n,k'}[b_n^\dagger-b_n,b_n] \nonumber \\
&=&b_n+W_h(\omega_n)g_{n,k'} . \label{eq:theta-bn-com}
\een
Therefore, 
\ben 
&&\sum _n \delta g_{n,jk} W_h(\omega_n) (\theta_{k'} b_n^\dagger \theta_{k'}^\dagger e^{i\omega_nt} - \theta_{k'} b_n \theta_{k'}^\dagger e^{-i\omega_nt} ) \nonumber \\
&&= \sum _n \delta g_{n,jk} W_h(\omega_n) (b_n^\dagger e^{i\omega_nt} - b_n e^{-i\omega_nt} ) \nonumber \\
&&+2i\sum_n \delta g_{n,jk} g_{n,k'}W_h(\omega_n)^2 \sin (\omega_n t) .
\een
Employing the above identity and using the definition of Eq. (\ref{eq:ft}), one can show that Eq. (\ref{eq:bcor1-app-b}) can be expressed as
\ben
&&\theta_{k'}\theta_j^\dagger(t)\theta_{k}(t)= \theta_{k'}\theta_j^\dagger(t)\theta_{k}(t)\theta_{k'}^\dagger\theta_{k'} \nonumber \\
&&=e^{\sum _n \delta g_{n,jk} W_h(\omega_n) (\theta_{k'} b_n^\dagger \theta_{k'}^\dagger e^{i\omega_nt} - \theta_{k'} b_n \theta_{k'}^\dagger e^{-i\omega_nt} )} \theta_{k'}\nonumber \\
&&=f_{jk,k'}(t)\theta_j^\dagger(t)\theta_{k}(t) \theta_{k'} .
\een 
The above identity implies that \re{the first term of Eq. (\ref{eq:b-deltarho}) (without $J_{jk}$)} can be simplified to
\ben
Tr_b \left \{ \theta_{k'} \theta_j^\dagger (t) \theta_k(t) \re{\theta_{j'}^\dagger} \rho_b \right\} &=& f_{jk,k'} (t)  \langle \theta_j^\dagger (t) \theta_k(t) \theta_{k'} \theta_{j'}^\dagger \rangle \nonumber \\
&=&f_{jk,k'} (t) w_{jk}w_{j'k'} e^{-{\mathcal K}_{jk,j'k'}(t)} , \nonumber \\\label{eq:ihom1-1}
\een
where Eq. (\ref{eq:theta-4}) has been used.

On the other hand, in the last term of Eq. (\ref{eq:b-deltarho}), product of the first two operators within the trace operation can be expressed as
\ben
&&\theta_{k'}D_j(t)= \theta_{k'}D_j(t)\theta_{k'}^\dagger \theta_{k'}\nonumber \\
&&=\sum_n \hbar\omega_n g_{n,j} (1-W_h(\omega_n)) \nonumber \\
&&\hspace{.5in}\times (\theta_{k'} b_n \theta_{k'}^\dagger e^{-i\omega_n t} +\theta_{k'} b_n^\dagger \theta_{k'}^\dagger e^{i\omega_nt})  \theta_{k'}  \nonumber \\
&&=D_j(t)\theta_{k'} +2\theta_{k'}\sum_n \hbar\omega_n g_{n,j}g_{n,k'} \nonumber \\ 
&&\hspace{.7in}\times (1-W_h(\omega_n)) W_h(\omega_n) \cos (\omega_n t) ,
\een
where Eqs. (\ref{eq:theta-bnd-com}) and (\ref{eq:theta-bn-com}) have been used.  The above identity leads to the following expression for the trace of the bath operators in the last term of Eq. (\ref{eq:b-deltarho}): 
\re{\be
Tr_b\left\{\theta_{k'} D_j(t) \theta_{j'}^\dagger \rho_{b} \right\} =\left ({\mathcal M}_{j,j'k'}(t) +h_{j,k'}(t) \right ) w_{j'k'}  , \label{eq:ihom1-2}
\ee
where the identity of }Eq. (\ref{eq:djt-1}) for $\tau=0$, the definition of Eq. (\ref{eq:mt}), and the definition of Eq. (\ref{eq:ht}) have been used.  Inserting Eqs. (\ref{eq:ihom1-1}) and (\ref{eq:ihom1-2}) into Eq. (\ref{eq:b-deltarho}), one can then obtain the following expression:
\ben
&&Tr_b\{ \tilde B_{jk} (t)\delta \tilde \rho_{b,j'k'} \}\nonumber \\
&&=J_{jk}w_{jk}(f_{jk,k'} (t) e^{-{\mathcal K}_{jk,j'k'}(t)}-1)w_{j'k'}  \nonumber \\
&& +\delta_{jk} ({\mathcal M}_{j,j'k'}(t) +h_{j,k'}(t)) w_{j'k'} . \label{eq:b-deltarho-1}
\een
It is easy to confirm that this expression is equivalent to Eq. (\ref{eq:ihom1-bath}).

\section{Evaluation of the bath correlation function in the second order inhomogeneous term}
The trace over the bath in the second order inhomogeneous term, Eq. (\ref{eq:ihom2}),  can be expressed  as follows:
\ben
&&Tr_b \left \{\tilde B_{jk}(t)\tilde B_{j'k'}(\tau) \delta \tilde \rho_{b,j''k''} \right\} \nonumber \\
&&=Tr_b \left \{\left (J_{jk} (\theta_j^\dagger (t)\theta_k(t)-w_{jk}) +\delta_{jk}D_j(t)\right )  \right . \nonumber \\
&&\times \left (J_{j'k'}(\theta_{j'}^\dagger(\tau)\theta_{k'}(\tau)-w_{j'k'} ) +\delta_{j'k'} D_{j'} (\tau) \right) \nonumber \\
&&\times \left . \left (\theta_{j''}^\dagger \rho_b \theta_{k''} -w_{j''k''} \rho_b \right) \right\} \nonumber \\
&&=J_{jk}J_{j'k'} F^{j''k''}_{jk,j'k'}(t,\tau) +\delta_{jk}J_{j'k'} H^{(1),j''k''}_{j,j'k'}(t,\tau) \nonumber \\
&&+J_{jk}\delta_{j'k'} \re{H^{(2),j''k''}_{jk,j'}(t,\tau)}+\delta_{jk}\delta_{j'k'}L^{j''k''}_{j,j'}(t,\tau) , \label{eq:inhom2-der}
\een
where
\ben
&&F^{j''k''}_{jk,j'k'}(t,\tau) \nonumber \\
&&= Tr_b\left \{ (\theta_j^\dagger (t)\theta_k(t)-w_{jk}) (\theta_{j'}^\dagger(\tau)\theta_{k'}(\tau)-w_{j'k'} ) \right . \nonumber \\
&&\hspace{1in}\left . \times (\theta_{j''}^\dagger \rho_b \theta_{k''} -w_{j''k''} \rho_b)\right\} ,  \label{eq:ihom2-c1}\\
&&H_{j,j'k'}^{(1),j''k''} (t,\tau)=Tr_b\left \{D_j(t) (\theta_{j'}^\dagger(\tau)\theta_{k'}(\tau)-w_{j'k'} )\right .\nonumber \\
&&\hspace{1in}\left .\times (\theta_{j''}^\dagger \rho_b \theta_{k''} -w_{j''k''} \rho_b)\right\} , \label{eq:ihom2-c2}\\
&&\re{H_{jk,j'}^{(2),j''k''} (t,\tau)}=Tr_b\left \{(\theta_{j}^\dagger(t)\theta_{k}(t)-w_{jk} )D_{j'}(\tau) \right .\nonumber \\
&&\hspace{1in}\left .\times (\theta_{j''}^\dagger \rho_b \theta_{k''} -w_{j''k''} \rho_b)\right\} , \label{eq:ihom2-c3}\\
&&L_{j,j'}^{j''k''} (t,\tau) =Tr_b\left \{D_j(t)D_{j'}(\tau) \right .\nonumber \\
&&\hspace{1in}\left . \times (\theta_{j''}^\dagger \rho_b \theta_{k''} -w_{j''k''} \rho_b)\right\} . \label{eq:ihom2-c4}
\een
Further calculation of each of the terms above is straightforward as described below.   

The first bath term in Eq. (\ref{eq:inhom2-der}), $F^{j''k''}_{jk,j'k'}(t,\tau)$ defined by Eq. (\ref{eq:ihom2-c1}), can be expanded further and expressed as follows:
\ben
&&F^{j''k''}_{jk,j'k'}(t,\tau) =\langle \theta_{k''}\theta_j^\dagger(t)\theta_k(t)\theta_{j'}^\dagger(\tau)\re{\theta_{k'}(\tau)}\theta_{j''} \rangle\nonumber \\
&&\hspace{.3in}-w_{jk} \langle \theta_{k''}\theta_{j'}^\dagger(\tau)\theta_{k'}(\tau)\theta_{j''}^\dagger \rangle \nonumber \\
&&\hspace{.3in}-w_{j'k'} \langle \theta_{k''}\theta_{j}^\dagger(t)\theta_{k}(t)\theta_{j''}^\dagger \rangle \nonumber \\
&&\hspace{.3in}+w_{jk}w_{j'k'} \langle \theta_{k''}\theta_{j''}^\dagger \rangle\nonumber \\
&&\hspace{.3in}-w_{j''k''} \langle (\theta_j^\dagger(t)\theta_k(t)-w_{jk})(\theta_{j'}^\dagger(\tau)\re{\theta_{k'}(\tau)}-w_{j'k'})\rangle . \nonumber \\ \label{eq:c1}
\een
The first term in the above expression can be calculated as follows:
\ben
&&\langle \theta_{k''}\theta_j^\dagger(t)\theta_k(t)\theta_{j'}^\dagger(\tau)\re{\theta_{k'}(\tau)}\theta_{j''} \rangle \nonumber \\
&&=f_{jk,k''}(t)f_{j'k',k''}(\tau) \langle \theta_j^\dagger(t)\theta_k(t)\theta_{j'}^\dagger(\tau)\re{\theta_{k'}(\tau)}\theta_{k''}\theta_{j''} \rangle \nonumber \\
&&=f_{jk,k''}(t)f_{jk',k''}(\tau)w_{jk}w_{j'k'}w_{j''k''} \nonumber \\
&&\hspace{.5in}\times e^{-{\mathcal K}_{jk,j''k''}(t)-{\mathcal K}_{j'k',j''k''}(\tau)-{\mathcal K}_{jk,j'k'}(t-\tau)} .
\een
For the second and third terms of Eq. (\ref{eq:c1}), Eq. (\ref{eq:ihom1-1}) can be used.   The last term of Eq. (\ref{eq:c1}) corresponds to the first term that appears in the evaluation of ${\mathcal R}(t)$.  Combining all of these, one can show that 
\ben
&&F^{j''k''}_{jk,j'k'}(t,\tau) =w_{jk}w_{j'k'}w_{j''k''} \left \{ e^{-{\mathcal K}_{jk,j'k'}(t-\tau)} \right . \nonumber\\
&&\ \times\left (f_{jk,k''}(t)f_{j'k',k''}(\tau) e^{-{\mathcal K}_{jk,j''k''}(t)-{\mathcal K}_{j'k',j''k''}(\tau)} -1\right ) \nonumber  \nonumber \\
&&\left .\ -f_{jk,k''}(t) e^{-{\mathcal K}_{jk,j''k''}(t)}-f_{j'k',k''}(\tau) e^{-{\mathcal K}_{j'k',j''k''}(\tau)}+2 \right \} .\nonumber\\
\een

\re{The bath term contributing to the second term of  Eq. (\ref{eq:inhom2-der}), {\it i.e.}, $H^{(1),j''k''}_{jk,j'k'}(t,\tau)$ defined by Eq. (\ref{eq:ihom2-c2}),} is expressed as follows:
\ben
&&H^{(1),j''k''}_{j,j'k'}(t,\tau) =\langle \theta_{k''}D_j(t)\theta_{j'}^\dagger(\tau)\theta_{k'}(\tau)\theta_{j''}^\dagger \rangle\nonumber \\
&&\hspace{.3in}-w_{j'k'} \langle \theta_{k''}D_j(t)\theta_{j''}^\dagger \rangle \nonumber \\
&&\hspace{.3in}-w_{j''k''} \langle D_j(t)(\theta_{j'}^\dagger(\tau)\theta_{k'}(\tau)-w_{j'k'})\rangle . \label{eq:c2}
\een
The first term in the above expression can be shown to be 
\ben
&&\langle \theta_{k''}D_j(t)\theta_{j'}^\dagger(\tau)\theta_{k'}(\tau)\theta_{j''}^\dagger \rangle \nonumber \\
&&=f_{j'k',k''}(\tau)\langle D_j(t)\theta_{j'}^\dagger(\tau)\theta_{k'}(\tau)\theta_{k''}\theta_{j''}^\dagger \rangle \nonumber \\
&&+f_{j'k',k''}(\tau)h_{j,k''}(t) \langle \theta_{j'}^\dagger(\tau)\theta_{k'}(\tau)\theta_{k''}\theta_{j''}^\dagger \rangle , \label{eq:c2-1}
\een
where the following identities have been used,
\ben 
&&\theta_{k''}D_j(t)\theta_{j'}^\dagger(\tau)\theta_{k'}(\tau)\nonumber \\
&&=\theta_{k''}D_j(t)\theta_{k''}^\dagger \theta_{k''} \theta_{j'}^\dagger(\tau)\theta_{k'}(\tau)\theta_{k''}^\dagger\theta_{k''} , \\
&&\theta_{k''}D_j(t)\theta_{k''}^\dagger=D_j(t)+h_{j,k''}(t) ,\label{eq:dj-disp}\\
&&\theta_{k''} \theta_{j'}^\dagger(\tau)\theta_{k'}(\tau)\theta_{k''}^\dagger=f_{j'k',k''}(\tau)\theta_{j'}^\dagger(\tau)\theta_{k'}(\tau) ,
\een 
along with the definitions of Eqs. (\ref{eq:ft}) and (\ref{eq:ht}). 
In Eq. (\ref{eq:c2-1}), the first bath average on the right hand side can be calculated employing identities similar to those leading to Eq. (\ref{eq:b3-1}).  The resulting expression is as follows:
\ben
&&\langle D_j(t)\theta_{j'}^\dagger(\tau)\theta_{k'}(\tau)\theta_{k''}\theta_{j''}^\dagger \rangle \nonumber \\
&&=({\mathcal M}_{j,j'k'}(t-\tau)+{\mathcal M}_{j,j''k''}(t) )w_{j'k'}w_{j''k''} e^{-{\mathcal K}_{j'k',j''k''}(\tau)} .\nonumber \\
\een
Combining this with the identity given by Eq. (\ref{eq:theta-4}) (with the replacement of  $t\rightarrow \tau $ and $\tau\rightarrow 0$), one can show that Eq. (\ref{eq:c2-1}) can be expressed as follows:
\ben
&&\langle \theta_{k''}D_j(t)\theta_{j'}^\dagger(\tau)\theta_{k'}(\tau)\theta_{j''}^\dagger \rangle \nonumber \\
&&=f_{j'k',k''}(\tau)\left ({\mathcal M}_{j,j'k'}(t-\tau)+{\mathcal M}_{j,j''k''}(t) +h_{j,k''}(t) \right) \nonumber \\
&&\hspace{.3in}\times w_{j'k'}w_{j''k''} e^{-{\mathcal K}_{j'k',j''k''}(\tau)} . \label{eq:c2-1-f}
\een

The second bath average term on the right hand side of Eq. (\ref{eq:c2}) can be shown to be 
\be
\langle \theta_{k''}D_j(t)\theta_{j''}^\dagger\rangle=({\mathcal M}_{j,j''k''}(t)+h_{j,k''}(t))w_{j''k''} , 
\ee
where Eqs. (\ref{eq:b3-1}) and (\ref{eq:dj-disp}) along with the definition of Eq. (\ref{eq:mt}) have been used.

For the last term on the righthand side of Eq. (\ref{eq:c2}), Eq. (\ref{eq:djt-1}) can be employed.   As a result, Eq. (\ref{eq:c2}) can be expressed as 
\ben
&&H^{(1),j''k''}_{j,j'k'}(t,\tau) =\left (f_{j'k',k''}(\tau)e^{-{\mathcal K}_{j'k',j''k''}(\tau)} -1\right )\nonumber \\
&&\times w_{j'k'}w_{j''k''} \left (M_{j,j'k'}(t-\tau)+M_{j,j''k''}(t)+h_{j,k''}(t) \right ) .  \nonumber \\  \label{eq:c2-2}
\een

\re{$H^{(2),j''k''}_{jk,j'}(t,\tau)$ defined by Eq. (\ref{eq:ihom2-c3}) can be calculated in a similar manner, but can in fact be calculated using its relation to $H^{(1),k''j''}_{j',kj}(\tau,t)$ as follows:    
\ben
&&H^{(2),j''k''}_{jk,j'}(t,\tau)=H^{(1),k''j''}_{j',kj}(\tau,t)^* \nonumber \\
&&=\left (f_{kj,j''}(t)e^{-{\mathcal K}_{kj,k''j''}(t)} -1\right )^*\nonumber \\
&&\times w_{jk}w_{j''k''} \left (M_{j',kj}(\tau-t)+M_{j',k''j''}(\tau)+h_{j',j''}(\tau) \right )^* \nonumber \\
&&=\left (f_{jk,k''}(t)e^{-{\mathcal K}_{jk,j''k''}(t)} -1\right )\nonumber \\
&&\times w_{jk}w_{j''k''} \left (M_{j',kj}(t-\tau)+M_{j',k''j''}(-\tau)+h_{j',j''}(\tau) \right ) .  \nonumber \\  \label{eq:c2-2}
\een
The first equality of the above equation can be confirmed from the definitions, Eqs. (\ref{eq:ihom2-c2}) and (\ref{eq:ihom2-c3}), and the third equality results from the following identities: $\left (f_{kj,j''}(t)e^{-{\mathcal K}_{kj,k''j''}(t) }\right)^*=f_{jk,k''}(t) e^{-{\mathcal K}_{jk,j''k''}(t)}$; $M_{j',kj}(\tau-t)^*=M_{j',kj}(t-\tau)$; $M_{j',k''j''}(\tau)^*=M_{j',k''j''}(-\tau)$; $h_{j',j''}(\tau)^*=h_{j',j''}(\tau)$.}  

Finally, Eq. (\ref{eq:ihom2-c4}) can be expressed as 
\ben
L^{j''k''}_{j,j'} (t,\tau) &=&\langle \theta_{k''}D_j(t)D_{j'}(\tau) \theta_{j''}^\dagger\rangle \nonumber \\ 
&&-w_{j''k''} \langle D_j(t) D_{j'}(\tau)\rangle  \ . \label{eq:l-jk}
\een
In the above expression, the first term on the right hand side can be expressed as
\ben
&&\langle \theta_{k''}D_{j}(t)D_{j'}(\tau) \theta_{j''}^\dagger \rangle =\langle D_j(t)D_{j'}(\tau)\theta_{k''}\theta_{j''}^\dagger \rangle \nonumber \\
&&\hspace{.2in}+h_{j,k''}(t)\langle D_{j'}(\tau) \theta_{k''}\theta_{j''}^\dagger \rangle \nonumber \\
&&\hspace{.2in}+h_{j',k''}(\tau) \langle D_j(t) \theta_{k''}\theta_{j''}^\dagger \rangle \nonumber \\
&&\hspace{.2in}+h_{j,k''}(t)h_{j',k''}(\tau) w_{j''k''}  , 
\een
where the first term can be calculated as follows:
\ben 
&&\langle D_j(t)D_{j'}(\tau)\theta_{k''}\theta_{j''}^\dagger \rho_b \rangle \nonumber \\
&&=\sum_n\sum_{n'} \hbar\omega_n g_{n,j}(1-W_h(\omega_n)) \hbar\omega_{n'} g_{n',j'} (1-W_h(\omega_{n'} ))  \nonumber \\
&&\hspace{.4in}\times \left \{ e^{-i(\omega_n t+\omega_{n'}\tau)} \langle b_nb_{n'}\theta_{k''}\theta_{j''}^\dagger \rangle \right . \nonumber \\
&&\hspace{.6in}+e^{-i(\omega_n t-\omega_{n'}\tau)} \langle b_n b_{n'}^\dagger \theta_{k''}\theta_{j''}^\dagger \rangle \nonumber \\ 
&&\hspace{.6in}+e^{i(\omega_n t-\omega_{n'}\tau)} \langle b_n^\dagger b_{n'}\theta_{k''}\theta_{j''}^\dagger \rangle \nonumber \\
&&\hspace{.6in}\left . +e^{-i(\omega_n t+\omega_{n'}\tau)} \langle b_n^\dagger b_{n'}^\dagger \theta_{k''}\theta_{j''}^\dagger \rangle \right \}  . \label{eq:c4-term1}
\een
In the above expression, the four averages over the bath can be calculated explicitly and can be expressed as
\ben
&&\langle b_n b_{n'} \theta_{k''}\theta_{j''}^\dagger \rangle =w_{j''k''} \frac{\delta g_{n,j''k''} W_h(\omega_n)}{(1-e^{-\beta\hbar\omega_n})}\nonumber \\
&&\hspace{.5in}\times \frac{\delta g_{n',j''k''} W_h(\omega_{n'})}{(1-e^{-\beta\hbar\omega_{n'}})} , \label{eq:bn-bn-th-kj} \\
&&\langle b_n b_{n'}^\dagger \theta_{k''}\theta_{j''}^\dagger \rangle =w_{j''k''}\left (\frac{1}{1-e^{-\beta\hbar\omega_n}}\delta_{nn'}\right .\nonumber \\
&&\left . -  \frac{\delta g_{n,j''k''} W_h(\omega_n)}{(1-e^{-\beta\hbar\omega_n})} \frac{\delta g_{n',j''k''} W_h(\omega_{n'})}{(1-e^{-\beta\hbar\omega_{n'}})}  e^{-\beta\hbar\omega_{n'}} \right ) , \label{eq:bn-bnd-th-kj} \\
&&\langle b_n^\dagger  b_{n'} \theta_{k''}\theta_{j''}^\dagger \rangle =w_{j''k''}\left ( \frac{e^{-\beta\hbar\omega_n}}{1-e^{-\beta\hbar\omega_n}}\delta_{nn'} \right .  \nonumber \\
&&\left . -\frac{\delta g_{n,j''k''} W_h(\omega_n)}{(1-e^{-\beta\hbar\omega_n})}e^{-\beta\hbar\omega_{n}} \frac{\delta g_{n',j''k''} W_h(\omega_{n'})}{(1-e^{-\beta\hbar\omega_{n'}})}  \right ),  \label{eq:bnd-bn-th-kj} \\
&&\langle b_n^\dagger b_{n'}^\dagger \theta_{k''}\theta_{j''}^\dagger \rangle =w_{j''k''} \frac{\delta g_{n,j''k''} W_h(\omega_n)}{(1-e^{-\beta\hbar\omega_n})} e^{-\beta\hbar\omega_{n}}\nonumber \\
&&\hspace{.5in}\times \frac{\delta g_{n',j''k''} W_h(\omega_{n'})}{(1-e^{-\beta\hbar\omega_{n'}})} e^{-\beta\hbar\omega_{n'}} .   \label{eq:bnd-bnd-th-kj}
\een
\re{It is worth noting that the additional factor for $n=n'$ in Eq. (\ref{eq:bn-bnd-th-kj})  comes from the first term of the following identity:
\ben
&&\langle b_n b_n^\dagger e^{\gamma^* b_n^\dagger -\gamma b_n}\rangle =\left ( \frac{1}{(1-e^{-\beta\hbar \omega_n} )} \right . \nonumber \\
&&\hspace{.2in}\left . -|\gamma|^2 \frac{e^{-\beta \hbar \omega_n}}{(1-e^{-\beta \hbar \omega_n})^2} \right ) e^{-\coth(\beta\hbar\omega_n/2)|\gamma|^2/2} 
\een 
For the case of (\ref{eq:bnd-bn-th-kj}), the fact that $b_n^\dagger b_n =b_n b_n^\dagger -1$ can be combined with the above identity.}
\re{When Eqs. (\ref{eq:bn-bn-th-kj})-(\ref{eq:bnd-bnd-th-kj}) are inserted into Eq. (\ref{eq:c4-term1}), the term for $n=n'$ cancel $-w_{j''k''} \langle D_j(t) D_{j'}(\tau)\rangle$ in Eq. (\ref{eq:l-jk}) and the double summation for $n$ and $n'$ can be factored as follows:}
\ben
\re{L^{j''k''}_{j,j'} (t,\tau)} &=&w_{j''k''} ({\mathcal M}_{j,j''k''}(t)+h_{j,k''}(t)) \nonumber \\ 
&& \times ({\mathcal M}_{j',j''k''}(\tau)+h_{j',k''}(\tau)) .
\een

For all the  bath correlation functions calculated above, let us define the following time integrals:
\ben
&&\tilde F_{jk,j'k'}^{j''k''}(t;{\mathcal E})=\int_0^t d\tau e^{i{\mathcal E} (t-\tau)/\hbar} F_{jk,j'k'}^{j''k''}(t,\tau) , \label{eq:til-f1}\\
&&\tilde H_{j,j'k'}^{(1),j''k''}(t;{\mathcal E})=\int_0^t d\tau e^{i{\mathcal E} (t-\tau)/\hbar} H_{j,j'k'}^{(1),j''k''}(t,\tau) , \nonumber \\ \label{eq:til-f2}\\
&&\tilde H_{jk,j'}^{(2),j''k''}(t;{\mathcal E})=\int_0^t d\tau e^{i{\mathcal E} (t-\tau)/\hbar} H_{jk,j'}^{(2),j''k''}(t,\tau) , \nonumber\\ \label{eq:til-f3}\\
&&\tilde L_{j,j'}^{j''k''}(t;{\mathcal E})=\int_0^t d\tau e^{i{\mathcal E} (t-\tau)/\hbar} L_{j,j'}^{j''k''}(t,\tau) . \label{eq:til-f4}  
\een
The above time correlation functions are useful for expressing the second order inhomogeneous terms in the basis of eigenstates of $\tilde H_s$. \\

\newpage
\noindent
{\bf REFERENCES}

\end{document}